\documentclass[a4paper,11pt]{article}
\usepackage{jheppub} 
\usepackage{lineno}

\usepackage{amsmath,mathrsfs,amssymb}
\usepackage{graphicx}
\usepackage{subfigure}
\usepackage{float}

\title{\boldmath Transit time of black holes on generalized free energy landscape}

\author[a]{Tianqi Yue}
\author[b,1]{Jin Wang}
\affiliation[a]{
 College of Physics,
  Jilin University,
  China}
\affiliation[b]{ Department of Chemistry, and Department of Physics and Astronomy,
  Stony Brook , USA}

\emailAdd{yuetq23@mails.jlu.edu.cn,\\
jin.wang.1@stonybrook.edu}
\abstract{Recently, the thermodynamics and kinetics of black hole phase transitions have garnered attention, particularly with the black hole radius being utilized as an order parameter within a generalized free energy landscape. In this framework, the local minima and maxima of the free energy correspond to stable and unstable states, respectively, while other states on the free energy landscape represent fluctuating black holes. Thermal fluctuations enable transitions between these stable states, and this stochastic kinetic behavior can be effectively described by the probabilistic Fokker-Planck equation.The transit time, defined as the time required for transitions or jumps between states, is a crucial physical quantity in phase transition kinetics, as it helps characterize the switching dynamics. By employing a harmonic transition state approximation, we examine the Hawking-Page phase transition and the Reissner-Nordström-Anti-de Sitter (RNAdS) black hole phase transition within the context of the generalized free energy landscape. We analytically quantify the transit time and its probability distribution for actual transition events, revealing the relationship between the mean transit time and the prefactor of the classical mean first passage time (MFPT). As the mean transit time decreases, the probability distribution narrows, indicating reduced fluctuations in transit time. These phenomena are connected to the topological structure of the generalized free energy landscape, the curvature of the free energy in both stable and unstable states, and the height of the energy barrier. We conclude that, for black hole phase transitions, the transit time serves as a characteristic timescale that reflects the actual jump time between states or to a transition state. Moreover, its relationship with the prefactor allows for a more precise quantification of the MFPT.}
\footnotetext[1]{Corresponding author:jin.wang.1@stonybrook.edu}

\begin{document}
\maketitle

\flushbottom

\section{Introduction}
Black holes are solutions to the Einstein field equations. Hawking's study of black hole collapse led to the discovery of Hawking radiation. Alongside the Unruh effect \cite{Hawking:1975vcx}, this highlighted the importance of the thermal nature of black holes. The laws of black hole thermodynamics have since been developed \cite{hawking1974}, including the well-known black hole area law. This law establishes a thermodynamic relationship between the surface gravity and the area of a black hole. Specifically, it states that the surface area of a black hole does not decrease but rather increases when an object falls into it. This \cite{Bekenstein:1973ur} implies that stable black holes have a specific temperature, known as Hawking temperature \cite{Hawking:1975vcx}. Additionally, the black hole area law indicates a close relationship between the event horizon area of a black hole and its entropy.

It is well established that temperature and entropy are crucial characteristics for describing a thermodynamic system. Therefore, the potential for black holes to possess entropy or temperature is an important consideration. Bekenstein \cite{Bekenstein:1973ur} developed the concept of black hole entropy based on the black hole area law, proposing that black holes should be viewed as thermodynamic systems with physical temperature and entropy. This groundbreaking discovery has spurred further exploration into black hole thermodynamics. Consequently, questions arise regarding the possible emergence of thermodynamic phases in black holes, the occurrence of phase transitions, and the dynamics involved in such transitions.

Hut \cite{P.Hut:1977} investigated phase transitions of charged black holes in flat spacetime. Hawking and Page \cite{Hawking:1982dh} treated black holes as states within a thermodynamic framework and revealed a first-order phase transition from thermal AdS space to large AdS black holes at a certain critical temperature. Subsequent studies on charged AdS black holes have highlighted significant similarities between charged AdS black holes and van der Waals liquids. Notably, the extended phase space thermodynamics of charged AdS black holes can be expressed by treating the cosmological constant as a thermodynamic pressure \cite{gunasekaran2012, kubiznak2012}. Since then, the concept of black holes as thermal entities has been extensively used to investigate their thermodynamic properties. The hypothesis of black hole molecules has been employed to study these properties \cite{2015Insight, 2019Repulsive}, and the thermodynamics and kinetics of black holes have garnered substantial attention.

Significantly, researchers have uncovered \cite{J.M.Maldacena:1999, Gubser:1998bc, Witten:1998qj} the anti-de Sitter/conformal field theory (AdS/CFT) correspondence, which has advanced the study of AdS black hole phase transitions. Utilizing the AdS/CFT correspondence, the Hawking-Page transition can be interpreted as a confinement/deconfinement transition \cite{Witten:1998zw} in quantum chromodynamics (QCD). The similarities between AdS black holes and van der Waals liquids have also been extensively explored \cite{Cai:2013qga, Wei:2012ui, Zou:2013owa, Rajagopal:2014ewa, Mo:2014qsa, Xu:2015rfa, Fernando:2016sps, Yazdikarimi:2019jux, Dehyadegari:2016nkd, Dayyani:2017fuz, Dehyadegari:2018pkb}.

The first passage time is a crucial quantity for measuring the kinetic rate of phase transitions. It reflects the time required for a phase transition to occur from one stable state to another, or the residence time in a stable state. Since the first passage time is a random variable influenced by environmental thermal fluctuations, studying the mean first passage time (MFPT) is essential, as it characterizes the timescale of the first occurrence of this random process. Previous studies have successfully used the black hole radius as an order parameter on the generalized free energy landscape to determine the MFPT for black hole phase transitions \cite{Li:2020khm, Li:2022gfe,2022Kinetics}.

To study black hole phase transitions on the generalized free energy landscape, the following points need to be clarified: First, black holes should be regarded as thermal entities. Second, on the generalized free energy landscape, the local minima and maxima of the free energy correspond to stable and unstable states, respectively, which are the on-shell solutions to the stationary Einstein field equations. Other states on the landscape represent fluctuating black hole states, which are off-shell and do not satisfy the stationary Einstein field equations.
 
\begin{figure}[ht]
    \centering
    \includegraphics[width=0.4\textwidth]{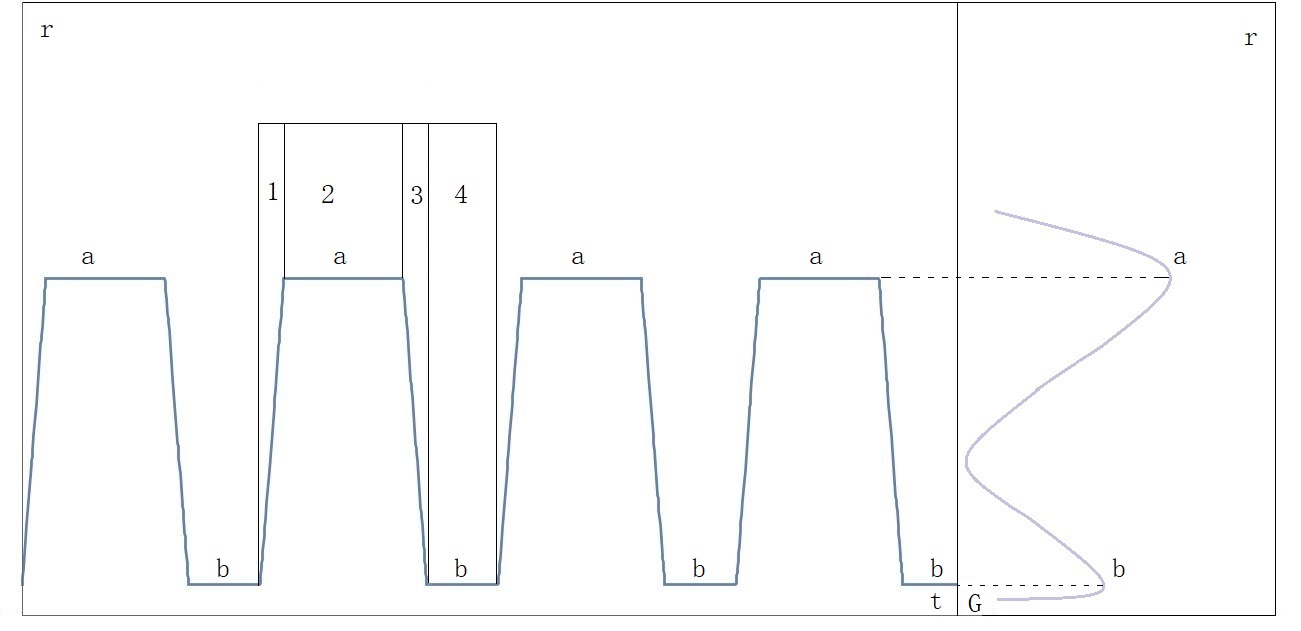}
    \caption{A schematic diagram of the phase transition time when two stable states undergo a phase transition.On the left side of the picture, the horizontal axis represents the time, the vertical axis represents the radius of the black hole, and on the right side the horizontal axis represents the free energy landscape and the vertical axis represents the radius of the black hole. 1 and 3 represent the transit time or jump time from one stable state a to b or vice versa. The mean first passage time from a to b or from b to a is the same as the residence time of staying in a or b states. }
    \label{Figure.syt}
\end{figure}
In this work, we introduce another crucial physical quantity for characterizing the kinetics of black hole phase transitions on the generalized free energy landscape, known as the transit time. It is important to emphasize that while both the mean first passage time (MFPT) and the mean transit time are characteristic timescales of black hole phase transitions, they represent different aspects. The MFPT quantifies the lifetime or residence time of a black hole in a stable state, whereas the mean transit time represents the actual duration required for a black hole to transition through a transition state and reach another stable state. These states are intuitively reflected in the free energy landscape through local maxima and minima.

Figure \ref{Figure.syt} illustrates four marked regions: 1, 2, 3, and 4. Region 1 represents the transit time required for a black hole to transition from a state with a smaller radius \( r \) to a state with a larger radius \( r \). Specifically, these two black hole states correspond to the minima of the free energy on the right side of the figure. Regions 2 and 4 denote the mean first passage time (MFPT) or residence time for large and small black holes, respectively. In these regions, the radius \( r \) remains constant over time, as indicated by the horizontal lines in the diagram. Finally, Region 3 depicts the transit time or jump time needed for a small black hole to transition into a large black hole. In summary, the transit time serves as a characteristic timescale that reflects the actual time taken to transition from one state to another or to a transition state. The mean first passage time, on the other hand, represents the time spent in a stable state or the waiting time for the stochastic switching to occur.

We aim to investigate the characteristics of the transit time and its role played in the black hole phase transition. The chapters of this paper are arranged as follows. In section \ref{section G} , we first construct the generalized free energy landscape using the Schwarzschild AdS black hole as an example. In  section \ref{section 3}, we introduce two quantities, mean first passage time and mean transit time, which are used to describe the dynamics of phase transitions. In section \ref{section 4}, we derive the mean transit time of black hole phase transition and its probability distribution. In section \ref{section 5}, the behaviour and physical significance between the generalized free energy landscape and the mean transit time and its probability distribution are discussed. In section \ref{section 6}, the discussion revolves around the dependence of transit time on temperature. 
 The section \ref{section 7},we summarize and discuss the results.

\section{Generalized free energy of the black hole}\label{section G}

In this section, we simplify the construction of the generalized free energy in black hole thermodynamics by leveraging the correspondence between black hole thermodynamics and classical thermodynamics.

For a more rigorous approach, the generalized free energy can be derived from the Einstein-Hilbert action of the Euclidean gravitational instanton with a conical singularity. This method accounts for the specific geometry of the gravitational field and provides a comprehensive and detailed understanding of the construction of the generalized free energy in black hole thermodynamics \cite{Li:2022gfe}.

Firstly, we consider the metric of a black hole, such as the Schwarzschild-AdS black hole, which is given by
\begin{equation}
   ds^{2}=-(1-\frac{2M}{r}+\frac{r^2}{L^2})dt^{2}+(1-\frac{2M}{r}+\frac{r^2}{L^2})^{-1}dr^{2}+r^{2}d\Omega^{2}     
\end{equation}
where \( M \) is the black hole mass, \( L = \sqrt{\frac{-3}{\Lambda}} \) is the curvature radius of the AdS black hole, and \( \Lambda \) is the cosmological constant.

One can derive various properties and quantities from the metric. Among these, the radius of the black hole event horizon is particularly important and can be obtained by
\begin{equation}
f(r)=1-\frac{2M}{r}+\frac{r^2}{L^2}=0
\end{equation}
so that we can determine the relationship between the black hole mass $M$, the AdS curvature radius $L$, and the event horizon radius $r_+$ in the Schwarzschild AdS black hole.
\begin{equation}
M=\frac{r_+}{2}\bigg(1+\frac{r_+^2}{L^2}\bigg)
\end{equation}

Black hole thermodynamics demonstrates a one-to-one correspondence between the mass and area of black holes and the fundamental quantities of classical thermodynamics, such as the relationship between entropy and the area of the horizon. Specifically, the Bekenstein-Hawking entropy is given by the area of the event horizon:
\begin{equation}
S = \frac{A}{4} =\pi r_+^2
\end{equation}

Here, \( A \) represents the black hole area. Additionally, the mass of a black hole corresponds to the internal energy \( U \) in thermodynamics and satisfies a similar relationship:
\begin{equation}\label{black hole t l}
dM=TdS+\Omega dJ +\Phi dQ
\end{equation}
Where \( \Omega \) represents the angular velocity, \( J \) denotes the angular momentum, \( \Phi \) signifies the electrostatic potential, and \( Q \) is the electric charge. In the Schwarzschild-AdS black hole, both the angular momentum \( J \) and the charge \( Q \) are zero, and we obtain the Hawking temperature \( T_H \) as follows:
\begin{equation}
T_H=\frac{\partial M}{\partial S}=\frac{\partial M}{\partial r_+}\frac{\partial r_+}{\partial S}=\frac{1}{4\pi r_+}\bigg(1+\frac{3r_+^2}{L^2}\bigg)
\end{equation}
Clearly, the minimum value of \( T_H \) can be obtained by taking the derivative of \( T_H \) with respect to \( r_+ \) and setting it to zero, which yields \( T_{\text{min}} = \frac{\sqrt{3}}{2 \pi L} \).

Building on the idea that there is a correspondence between black holes and thermodynamics, we can also determine the free energy of black holes.
\begin{equation}
G=M-T_H S=\frac{r_+}{2}\bigg(1+\frac{r_+^2}{L^2}\bigg)-\pi T_H r_+^2
\end{equation}
This series of operations reveals a clear picture. As the temperature varies from \( T_{\text{min}} \) to infinity, the geometry of the free energy landscape \( G \) changes accordingly. This changing geometry indicates that the system is locally stable only when the black hole radius \( r \) takes on certain values. In other words, the black hole radius is closely related to the temperature, with different temperature values corresponding to different geometric configurations in the free energy landscape.
\begin{figure}[!ht]
    \centering
    \includegraphics[width=0.4\textwidth]{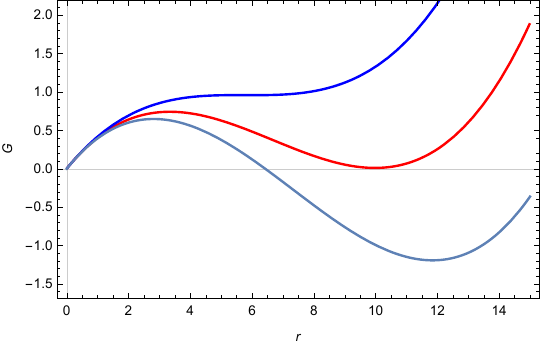}
    \caption{The generalized free energy landscapes at three different temperatures}
    \label{Figure.1}
\end{figure}
At this point, we consider the black hole radius as the order parameter and form an ensemble of black holes with varying event horizons at a specific Hawking temperature. We then replace the Hawking temperature with the ensemble temperature to obtain the generalized free energy landscape, which accounts for both stable and unstable states as well as fluctuating black holes within the ensemble.

As shown in the figure.\ref{Figure.1},we obtain the generalized free energy landscape.
\begin{equation}
G=\frac{r}{2}\bigg(1+\frac{r^2}{L^2}\bigg)-\pi T r^2
\end{equation}
In this framework, we refer to local maxima and minima as unstable and stable states, respectively, which are the on-shell solutions to the stationary Einstein field equations \cite{Li:2022gfe}. Conversely, black holes corresponding to different free energies are categorized as fluctuating states. These states can be understood as transient black hole configurations arising from thermal fluctuations. By incorporating these fluctuating states, we construct the generalized free energy landscape for black holes. In the subsequent chapter, we will define the transit time for kinetics on the free energy landscape more precisely, and derive both its probability distribution and its average value.

\section{From the generalized free energy landscape to kinetic characterization}\label{section 3}
In this section, we investigate the dynamics of black hole phase transitions or state switching. As illustrated in the previous section, the local maxima and minima of the generalized free energy landscape correspond to unstable and stable states, respectively. Due to environmental thermal fluctuations, each stable state has the potential to transition to other stable states, while unstable states serve as transition states, facilitating these phase transitions. Key quantities for characterizing this dynamic process are the mean first passage time (MFPT) and the mean transit time.

\subsection{The mean first passage time}
In the previous chapter, when the temperature exceeds \( T_{\text{min}} \), the free energy diagram as a function of the black hole radius displays a double-basin structure. Consequently, it is natural to use the black hole radius as the reaction coordinate or order parameter. The dynamics underlying these transitions can be described using the stochastic Langevin equation.

\begin{equation}
    m \ddot{r}=-m \eta \dot{r}-G'(r)+f(t)
    \label{LGE}
\end{equation}
Here, \( f(t) \) represents the random force and satisfies the fluctuation-dissipation relation:
\begin{equation}
    \big<f(t)f(t')\big> = 2k_{B}T\eta m \delta(t-t')
\end{equation}
where \( \eta \) is the friction coefficient. To further simplify the analysis, we assume the system is overdamped, which allows us to neglect the left-hand side of Equation \eqref{LGE}. Additionally, without loss of generality, we set the mass \( m \) to unity.

The Schwarzschild black hole phase, the thermal AdS phase, and any intermediate states during the Hawking-Page phase transition represent specific states, or groups of states, within the broader thermodynamic ensemble. Consequently, the stochastic kinetics of these states, influenced by thermal fluctuations, can be effectively described using the probabilistic Fokker-Planck equation.
\begin{equation}
\frac{\partial \rho(r,t)}{\partial t} = D \frac{\partial}{\partial r}\left( e^{-\beta G(r)}  \frac{\partial}{\partial r} (e^{\beta G(r)} \rho(r,t)) \right)
\end{equation}

In the aforementioned equation, \( \rho \) represents the probability density, \( \beta \) denotes the inverse temperature \( \frac{1}{k_B T} \), and \( D \) is the diffusion coefficient, given by \( \frac{k_B T}{\eta} \), where \( k_B \) is the Boltzmann constant and \( \eta \) is the dissipation or friction coefficient. To solve this equation and obtain the mean first passage time, the following boundary conditions can be imposed:

1. Reflecting Boundary Condition at $r_A$ is:
\begin{equation}
j(r_A,t) = -D  \left(e^{-\beta G(r)}  \frac{\partial}{\partial r} (e^{\beta G(r)} \rho(r,t))\right) \bigg|_{r=r_A} = 0
\end{equation}

This condition specifies that the probability flux \( j \) at the boundary \( r_A \) is zero, indicating a reflecting boundary where no particles flow out of the system at \( r_A \). Here, \( r_A \) is typically chosen as the initial state for the kinetics of the phase transition.

2. Absorbing Boundary Condition at $r_m$ is:
\begin{equation}
\rho(r_m,t) = 0
\end{equation}

This condition states that the probability density \( \rho \) at the boundary \( r_m \) is zero, indicating an absorbing boundary where particles are absorbed and the phase transition is considered complete. Note that \( r_m \) represents the value of \( r \) at which \( G = G_m \), corresponding to the maximum value of \( G \) within the range from \( r_A \) to \( r_B \).

To proceed, the mean first passage time can be obtained through the following steps.

In this context, the probability density \( \rho(r, t) \) represents the probability that the black hole is at position \( r \) at time \( t \). Given that \( r_m \) is chosen as the intermediate state boundary condition, the first step is to determine its probability distribution to calculate the mean first passage time.

We denote the distribution of first passage times by \( F_p(t) \),
\begin{equation}
    F_p(t)=-\frac{d( \int_{r_{A}}^{r_{m}} dr \rho)}{dt}
\end{equation}

It is evident that \( F_p \, dt \) represents the probability that the black hole transitions from the initial state to the intermediate state \( r_m \) within the time interval \( (t, t + dt) \). This implies that the change in survival probability is equal to the negative of the change in the distribution of kinetics for the escape. In other words, a higher chance of escape corresponds to a lower chance of survival. This allows us to establish the connection between the mean first passage time and \( \rho \), given by the equation:
\begin{equation}
    \big<t_{mfp}\big>=\int_{0}^{\infty}dt t F_p(t)
\end{equation}

Furthermore, the probability density \( \rho \) can be determined by solving the Fokker-Planck equation with the specified initial and boundary conditions. Consequently, we obtain:
\begin{equation}
\label{MFPT}
   \big<t_{mfp}\big>=\frac{1}{D}\int_{r_{A}}^{r_{m}}dr \int_{r_{A}}^{r}dr' e^{\beta(G(r)-G(r'))} 
\end{equation}

In reaction theory, the mean first passage time (MFPT) is often approximated by treating the barrier as a superposition of two quadratic potential energies: one at the local stable state (the minimum of free energy) and one at the local unstable state (the local maximum of free energy). For example, the free energy landscape can be locally approximated near the local minimum and local maximum as:
\begin{equation}\label{G js 1}
    G(r) \approx G(r_A) + \frac{1}{2}\omega_{A}^2(r-r_{A})^2 
\end{equation}  
\begin{equation}\label{G js 2}
    G(r) \approx G(r_m)- \frac{1}{2}\omega_{max}^2(r-r_{m})^2
\end{equation}
In this context, we employ two quadratic functions to approximate the generalized free energy landscape. In this work \cite{zwanzig2001nonequilibrium}, the mean first passage time (MFPT) can be expressed in the following form:

\begin{equation}
    \label{tT}
    \big<t_{mfp}\big> = \frac{2\pi\eta}{\omega_{A}\omega_{max}} e^{\beta \Delta G}
\end{equation}
In equations \eqref{G js 1}, \eqref{G js 2}, and \eqref{tT}, \(\omega_{A}\) and \(\omega_{max}\) represent the curvatures of the free energy \(A\) at the stable black hole state and \(r_m\) at the top of the barrier, respectively. The factor preceding the exponent is referred to as the prefactor of the mean first passage time (MFPT). It is clearly determined by the curvatures and the friction coefficient \(\eta\).

\subsection{The mean transit time}
As predicted by transition state theory, the transit time, as a characteristic timescale, reflects the actual time required for a system to transition from one state to another or to a transition state. This is analogous to the mean first passage time, which represents the time spent in a stable state or the waiting time for stochastic switching to occur (refer to Figure \ref{Figure.syt}). Consequently, the properties of the mean transit time are significantly influenced by the intermediate states. On the generalized free energy landscape, this influence is evident from the fact that the original free energy landscape can be approximated by a function that is expanded around the intermediate state.

\begin{equation}
    G(r) \approx G^{\neq} - \frac{1}{2}\omega_{max}^2(r-r_{m})^2
\end{equation}

Where \( G^{\neq} \) represents the free energy at the top of the barrier or transition state. Szabo derived an analytical expression for the mean transit time \cite{Makarov:2010}.
\begin{equation}
   \big<t_{TP}\big>= \frac{\eta ln(2e^{\gamma} \beta G^{\neq} )}{\omega^2_{max}}
   \label{ttp}
\end{equation}
where \(\gamma = 0.577\ldots\) is Euler's constant. According to the formula \eqref{ttp}, the transit time is logarithmically related to the height of the barrier. This implies that the height of the barrier has a relatively minor effect. However, the formula also indicates that the transit time is primarily determined by the curvature factor \(\omega_{\text{max}}\) at the transition state.

In addition, there is a special case where the curvatures of the stable state and the transition state are nearly the same, satisfying \(\omega_A \approx \omega_{\text{max}}\). In this scenario, the prefactor of the MFPT is given by \(\tau_0 = \frac{2\pi \eta}{\omega_A \omega_{\text{max}}}\). Interestingly, the mean transit time and the prefactor of the MFPT are then related as follows:
\begin{equation}
   \big<t_{TP}\big>\approx  \frac{\tau_0}{2\pi}ln(2e^{\gamma} \beta G^{\neq} )
\end{equation}

Even under more general conditions where the curvature does not exhibit the aforementioned relationship, it is noteworthy that the prefactor of the MFPT still exhibits trends similar to those observed for the mean transit time.
\begin{equation}
   \big<t_{TP}\big>=  \frac{\omega_A}{\omega_{max}}\frac{\tau_0}{2\pi}ln(2e^{\gamma} \beta G^{\neq} )
\end{equation}
Indeed, the close relationship between the prefactor of the MFPT and the mean transit time on the free energy landscape aligns well with our physical intuition. The dynamics of a system undergoing phase transitions are intricately linked to the shape and structure of the free energy landscape.

The above discussion is generally applicable, and in the subsequent chapter, we will apply these concepts to study black hole phase transitions.

\section{The transit time and its probability distribution for black hole}
\label{section 4}
In this section, we derive the mean transit time and its probability distribution for both the Hawking-Page phase transition and the RNAdS black hole phase transition. From the previous section, we learned that combining the Fokker-Planck equation with the topological structure of the generalized free energy reveals how the probability $\rho$ of a black hole with radius $r$ evolves over time $t$. Additionally, we discovered that the kinetic quantity mean first passage time (MFPT) corresponds to the time spent in a stable state or the waiting time for stochastic switching to occur.

Similarly, the transit time is crucial in the phase transition of black holes, especially when intermediate states are involved. On the free energy landscape of the Hawking-Page phase transition, a local maximum at the top of the free energy barrier indicates that the black hole at a radius of $r_m$ is in an unstable state. Intuitively, the transit time reflects the actual jump time from one state to another or to a transition state (see Figure \ref{Figure.syt}). By combining the two kinetic quantities—mean first passage time (MFPT) and mean transit time—one can provide a comprehensive description and uncover the underlying details of black hole phase transitions.

To describe this process, we consider the problem of one-dimensional crossing of a free energy barrier. The distribution of transit times can be derived using the method outlined in reference \cite{Makarov:2010}. The specific derivation method is detailed in Appendix \ref{appendix A}. For the Hawking-Page phase transition, the probability distribution of the transit time can be derived as follows:
\begin{equation}
P_{AB}(t)=2\omega\beta G^{\neq}e^{-\frac{\beta G^{\neq}}{sinh^2(\omega t)}}\frac{cosh(\omega t)}{sinh^3(\omega t)}
\end{equation}
where \( \omega \) represents the curvature at the top of the potential barrier in the free energy landscape. To estimate the mean transit time in the long-time asymptotic limit, \( P_{AB} \) becomes:
\begin{equation}\label{PAB Esh}
P_{AB}(t)=8\omega \beta G^{\neq} e^{-2\omega t -4\beta G^{\neq} e^{-2\omega t}}
\end{equation}
By applying the maximum likelihood method, we can derive the probability to obtain the following equation: 
\begin{equation}\label{tab Esh}
\big<t_{AB}\big>=\frac{ln[4\beta G^{\neq}]}{2\omega}
\end{equation}

The equation \eqref{tab Esh} represents a simplified model. For more general scenarios, the stochastic Langevin equation \eqref{LGE} should be employed to account for the effects of stochastic forces and friction on the transit time.

Applying the framework of harmonic transition state theory \cite{1985On,Pollak1986Theory,Makarov:2010}, we derive the mean transit time under overdamped conditions governed by the stochastic Langevin equation. In cases where the system is coupled to baths, the Hamiltonian can be expressed as:
\begin{equation}
    H=\frac{p^2}{2m}+G(r)+\sum_{i=1}^{n}\frac{p_{i}^2}{2}+\frac{\omega_i^2}{2}(r_i-\frac{c_i r}{\omega_i^2})^2
\end{equation}
Clearly, this Hamiltonian describes a system coupled with multiple harmonic oscillator baths. In simple terms, the harmonic oscillator variables \( r_i \) with coupling constants \( c_i \) to the black hole through the horizon size \( r \) and frequencies \( \omega_i \) characterize the friction and the random force \( f(t) \). By diagonalizing the system, these coupled harmonic oscillators can be decomposed into several normal modes. One of these modes has an imaginary frequency \( i\omega_K \), given by:
\begin{equation}
    \omega_K = \sqrt{\left(\frac{\eta^2}{4}+\omega^2\right)} - \frac{\eta}{2}
\end{equation}
Here, \( \omega \) continues to denote the curvature at the top of the potential barrier in the free energy landscape, which originates from approximating the free energy landscape around the transition state as a parabola. Under the overdamped condition where \( \eta \gg \omega \), it follows that:
\begin{equation}
    \omega_K \approx \frac{\omega^2}{\eta}
\end{equation}
Following Zwanzig et al.\ \cite{zwanzig2001nonequilibrium}, one can substitute \( \omega \) with \( \omega_K \) in equations \eqref{PAB Esh} and \eqref{tab Esh}. This substitution allows us to derive the probability distribution \( P_{AB} \) and the mean transit time for the black hole phase transition.

As an example, the probability distribution of the transit time for the transition of a Schwarzschild-AdS black hole, under the harmonic approximation of the free energy near the transition state, becomes:
\begin{equation}
   P_{AB}(t)=2\omega_K\beta G^{\neq} e^{-\frac{\beta G^{\neq}}{sinh^2(\omega_K t)}}\frac{cosh(\omega_K t)}{sinh^3(\omega_K t)}
   \label{PAB HP}
\end{equation}
Then the mean transit time becomes,
\begin{equation}\label{HPtransit time}
    \big<t_{AB}\big>=\frac{ln[4\beta G^{\neq}]}{2\omega_K}=\frac{\eta ln[4\beta G^{\neq}]}{2\omega^2}
\end{equation}
Thus, the mean transit time of the Hawking-Page phase transition under overdamped conditions is obtained. Compared to equation \eqref{ttp}, they are very similar; both have numerators related to the height of the potential barrier \( G^{\neq} \) with a logarithmic relationship, although the coefficients differ somewhat. Most importantly, there is an additional \( \frac{1}{2} \) factor in the denominator. This behavior can be well explained. By examining the free energy landscape of the Hawking-Page phase transition, when the temperature is above the critical temperature, the absolute difference between the AdS plasma phase and the intermediate transition black hole radius \( r_m - r_A \) is less than the absolute difference between the large black hole radius and the intermediate black hole radius \( r_B - r_m \). Moreover, this difference increases rapidly with rising temperature. In summary, the origin of the \( \frac{1}{2} \) factor is that we approximate the potential energy function as nearly half a parabola, while equation \eqref{ttp} assumed a symmetric parabola, considering the doubled distance traveled.

The analytical expression for the probability distribution of transition times for RNAdS black hole phase transitions can be derived. Details are provided in Appendix \ref{appendix B}.
\begin{equation}
   P_{AB}(t)=2\omega'_K\beta G^{\neq} e^{-\frac{\beta G^{\neq}}{sinh^2(\omega'_K t)}}\frac{cosh(\omega'_K t)}{sinh^3(\omega'_K t)} 
   \label{PAB RN}
\end{equation}
\begin{equation}\label{RN tranit time jx}
    \big<t_{AB}\big>=\frac{ln[4\beta G^{\neq}]}{2\omega'_K}
\end{equation}
\begin{equation}
    \omega'_K=\frac{\omega'^2}{\eta}
\end{equation}
where $\omega'$ denotes the curvature derived from formula \eqref{w'} in the appendix, obtained by approximating the free energy landscape of the RNAdS black hole as a parabola at the top of the barrier. Thus, the mean and distribution of the transit time for both the Hawking-Page phase transition and the RNAdS black hole phase transition under overdamped conditions are determined.

\section{ Behaviors and physical significance of the mean transit time for black hole}
\label{section 5}
In this section, we present the behavior of the mean transit time and its probability distribution. Key properties of the mean transit time for black holes are discussed.

It is important to note that black hole phase transitions involve an intermediate transition state between two stable states, as depicted in the generalized free energy landscape. This landscape features two basins separated by a barrier. 

The rate of phase transitions is determined by the curvature and height of the barriers in the Gibbs free energy landscape, which are influenced by factors such as temperature, charge, and the AdS curvature radius (or cosmological constant). To investigate the behavior of the transit time in black hole phase transitions, one must analyze the relationship between these parameters and the curvature and height of the barrier.

\subsection{Local generalized free energy landscape dictates mean transit time}
In this subsection, we discuss the analytical expression for the mean transit time. It becomes evident that the mean transit time is intricately linked to the geometric quantities of the generalized free energy landscape. We will highlight the physical significance of this relationship and demonstrate that using this analytical expression to investigate the properties of black hole phase transitions is both convenient and straightforward.

Based on the analytical expressions \eqref{HPtransit time} and \eqref{RN tranit time jx}, the mean transit time for both the Hawking-Page phase transition and the RNAdS black hole phase transition exhibits a similar structure.

The barrier height demonstrates a logarithmic dependence on the mean transit time and is inversely proportional to the square of the local curvature of the free energy landscape. The complex interactions between the black hole and its thermal environment, which drive black hole phase transitions, become evident through the construction of the generalized free energy. Just as Bekenstein's black hole entropy encodes one bit of information about the internal state of the black hole per fundamental unit of horizon area, the generalized free energy landscape captures information about the black hole phase transition. The local geometric features of this landscape are influenced by temperature and the AdS curvature radius or cosmological constant.
From another perspective, understanding the behavior of the transit time during black hole phase transitions provides valuable insights into the black hole itself. The temperature of the black hole and the AdS curvature radius or cosmological constant directly affect the curvature and barrier height of the free energy landscape, thereby influencing the transit time. In other words, the behavior of the transit time encodes information about the black hole.

Specifically, the curvature of the basins indicates the stability of the black hole: a steeper curvature reflects greater instability. The barrier height measures the difficulty for the black hole to transition through the barrier, while the curvature at the top of the barrier signifies the stability of the intermediate state of the black hole. The impact of the thermal environment on black hole phase transitions is evident in how varying temperatures alter the generalized free energy landscape. Additionally, the interactions between the black hole and its thermal environment are characterized by the friction coefficient \(\eta\).

When damping and random forces are not considered, the logarithmic dependence on the barrier height becomes more transparent. In this scenario, we only need to examine the particle at the top of the barrier using a harmonic approximation. Due to thermal fluctuations, the particle gains an initial velocity \( \frac{1}{2}m v_0^2 = \frac{1}{2} k_B T \), causing it to start moving from the top of the barrier. Newton's equation of motion \( m \ddot{x} = m \omega^2 x \) yields the general solution: \( x(t) = \frac{v_0}{\omega} \sinh(\omega t) \). The time spent passing through the barrier region between \( -x_0 \) and \( x_0 \) is then given by: \(t = \frac{2}{\omega} \sinh^{-1} \left( \frac{\omega x_0}{v_0} \right) \approx \frac{\ln(\beta \Delta V)}{\omega} \) 
where \( \Delta V = \frac{1}{2}\omega^2 x_0^2 \). Thus, the logarithmic dependence on the barrier height arises because the particle, starting from the top of the barrier (an unstable state), is influenced by thermal fluctuations, causing it to move toward other basins. The form of the mean transit time equations \eqref{HPtransit time} and \eqref{RN tranit time jx} takes into account these factors.

Although we approximate the generalized free energy landscape locally as a parabola, this approximation is relatively accurate because the time a black hole spends in stable and unstable states is much longer than its fluctuation time. Consequently, the local maxima and minima of the generalized free energy landscape play a crucial role in phase transitions. Analogous to classical physics, an object initially in the basin of a potential energy function receives random energy due to thermal fluctuations, gradually performing irregular movements until it acquires sufficient energy to jump over the barrier and stabilize in the next basin, resulting in a phase transition.

This behavior also explains why the mean transit time is related to the curvature at the top of the barrier. By using the generalized free energy landscape, the kinetic behavior of black hole phase transitions can be fully connected to the motion of classical particles in a potential field. Since the time a particle spends in stable and unstable states is much longer than in transient states, or considering that a particle begins to move randomly from an equilibrium state due to thermal fluctuations, non-equilibrium statistical transient processes must be considered. The time for this transient process is often much shorter than that of the equilibrium state. From the perspective of the free energy landscape, most of the particle's motion occurs near the basin and the top of the barrier. Therefore, after statistical averaging, the result heavily depends on the local quantity, which is the curvature at the top of the barrier.

Of course, there is additional information about phase transitions that the generalized free energy landscape cannot provide. For instance, the friction coefficient \( \eta \) reflects the complex interactions between the black hole and its environment. However, the specific value of this quantity cannot be directly extracted from the generalized free energy landscape.

With a deeper understanding of the mean transit time now achieved, the next step is to analyze its behavior by examining how various physical parameters affect the curvature and barrier height of the black hole's free energy landscape.

\subsection{Curvature and barrier height: shaping the probability distributions of the transit time}
In this subsection, we explore the relationship between the probability distribution and the generalized free energy landscape. By analyzing the probability distributions for the Hawking-Page phase transition (Equation \eqref{PAB HP}) and the RNAdS black hole phase transition (Equation \eqref{PAB RN}), and focusing specifically on the time-dependent terms, it is evident that both share a similar formal structure. This similarity highlights their dependence on the curvature of the potential barrier, reflecting how the geometry of the free energy landscape influences transit time probabilities during phase transitions.

In this context, the curvature resembles the oscillation frequency of a vibrating system. By analogy, a particle's random motion within a potential field, influenced by environmental thermal fluctuations, clarifies this behavior. A particle oscillates in a stable basin due to these fluctuations until it surpasses the potential barrier. This accumulation phase corresponds to the mean first passage time (MFPT). The transit time refers to the duration it takes for the particle to cross the barrier from one basin to another, with the majority of this time spent at the top of the barrier. A detailed examination suggests that transit events predominantly occur at the barrier's summit, manifesting as a characteristic frequency in the probability distribution. The stability of this intermediate state governs the process of transit events. Statistical aggregation of these events yields the analytical expressions given in Equations \eqref{PAB HP} and \eqref{PAB RN}. Here, the friction coefficient \( \eta \) is intricately linked with the curvature, together influencing the transit event dynamics.

The barrier height reflects the duration a black hole spends in a transient state, elucidating the logarithmic relationship between the barrier height and the mean transit time. This suggests a weaker correlation between these two quantities.

We now have a clearer understanding of how curvature, barrier height, and the friction coefficient \( \eta \) synergistically influence the transit event and its probability distribution. In the forthcoming chapters, we will explore in detail how these geometric features of the generalized free energy landscape affect the probability distribution, including their impact on the distribution's geometric form and its fluctuations.

\subsection{Mean transit time quantifies kinetic prefactor}
\subsubsection{Hawking-Page phase transition}
A comprehensive understanding of a system's phase transition time requires considering both the mean first passage time (MFPT) and the mean transit time. These temporal metrics are influenced by the free energy landscape, which includes the barrier height \(\Delta G\) or \(G^{\neq}\) and the curvature \(\omega\) of the barrier or basin. Notably, there is a significant correlation between the MFPT prefactor, which is affected by curvature, and the mean transit time.

To simplify the analysis, we use the extreme value approximation. This technique approximates the integrand of the integral \eqref{MFPT} by its maximum value within the integration domain, treating it as a constant for the purpose of integration. This method facilitates the straightforward evaluation of the integral.
\begin{equation}\label{MFPTjs}
    \frac{(r_m-r_A)^2}{2D} e^{\beta(G(r_m)-G(r_A))}
\end{equation}
\begin{figure}[ht]
    \centering

    \subfigure[]{
    \includegraphics[width=0.4\textwidth]{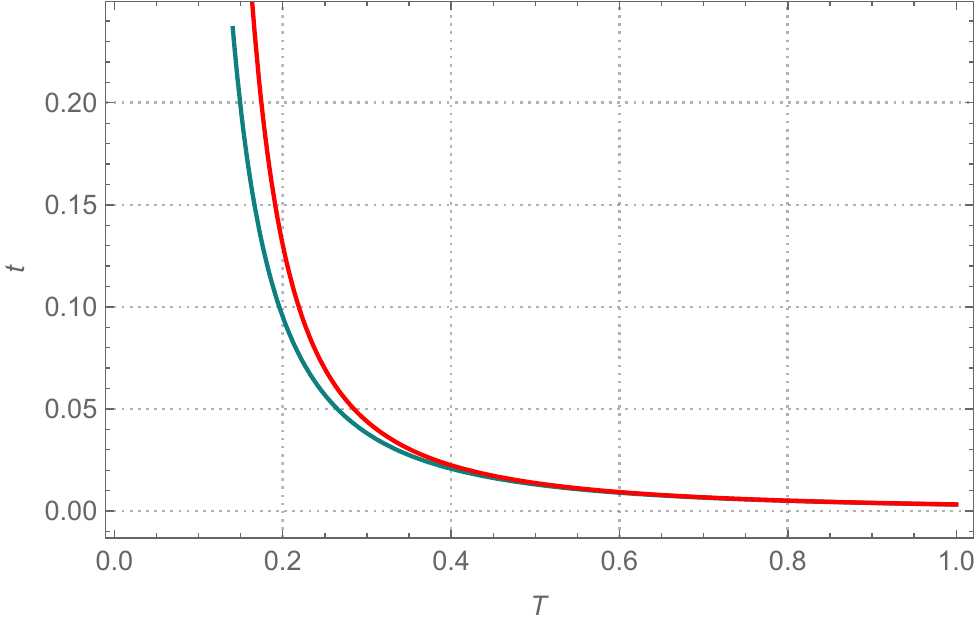}
    }
    \subfigure[]{
    \includegraphics[width=0.4\textwidth]{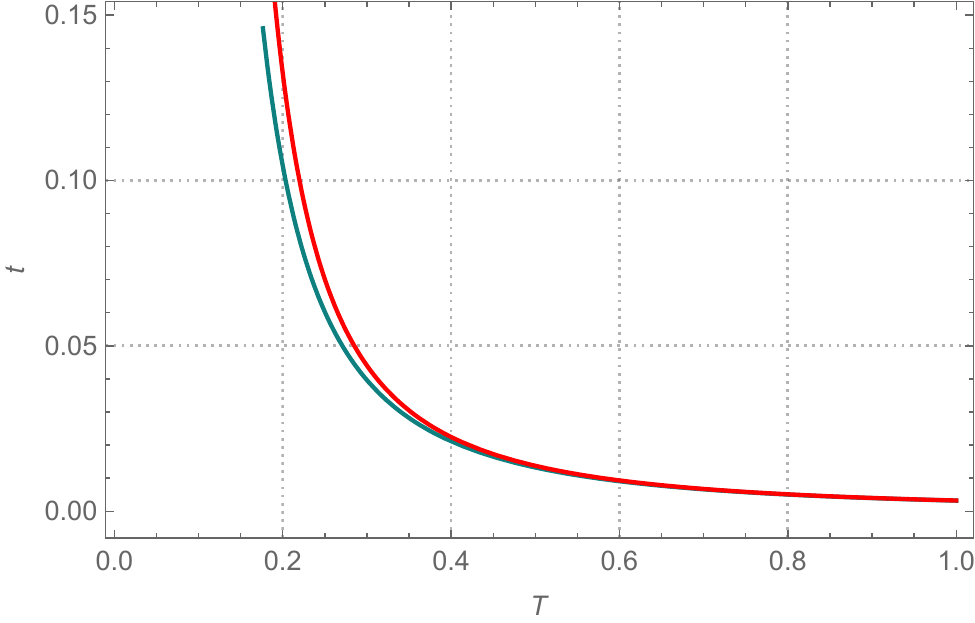}
    }
    
    \caption{Figure comparing the approximate results of MFPT for the Schwarzschild AdS black hole with the numerical results, where the red lines represent the approximate results and the green lines represent the numerical results. (a)$L^2$=100 (b)$L^2$=10}
    \label{fig:2mfpt}
\end{figure}
Based on Figure \ref{fig:2mfpt}, it is evident that the approximated results are quite reliable. Moreover, the error between the approximate results and the numerical results decreases as the temperature increases.

For the prefactor of the MFPT derived from \eqref{MFPTjs}, the following formula for a black hole phase transition can be obtained:
\begin{equation}\label{MFPTPrefactor}
  \tau_{0} = \frac{2\pi\eta}{\omega_{A}\omega_{max}}=\frac{(r_m-r_A)^2}{2D}
\end{equation}
By incorporating the condition \( r_A = 0 \) into equation \eqref{MFPTjs} and applying a Taylor series expansion to the term \( T_{min}^2/T^2 \) in equation \eqref{eq:rm} for small values, and considering \( D = \frac{k_B T}{\eta} \) with \( k_B = 1 \), one obtains:
\begin{equation}\label{HPmfpt}
   \tau_{0}= \frac{\eta}{32 \pi^2 T^3}
\end{equation}

To proceed, recalling equation \eqref{HPtransit time} and considering the relationship between the prefactor and the mean transit time, as well as the fact that \(\omega\) represents the curvature at the top of the barrier (\(\omega_{max}\)), it is evident that the mean transit time for black hole phase transitions should follow a similar pattern.
\begin{equation}\label{HPprefactor transit time}
    \big<t_{AB}\big>=\frac{\omega_{A}}{\omega}\frac{\tau_{0} }{4 \pi } ln[4\beta G^{\neq}]
\end{equation}

Given the assumption that \(\ln(4\beta G^{\neq})\) varies slowly, the mean transit time \(\big<t_{AB}\big>\) is proportional to the inverse of temperature \(T\), as expressed by the formula \(\big<t_{AB}\big> \propto \frac{\eta}{T}\). Examining equation \eqref{HPmfpt}, it is evident that, under the condition where \(T\) significantly exceeds the minimum temperature \(T_{min}\), the mean first passage time \(\big<t_{mfp}\big>\) for the Hawking-Page phase transition follows a different relationship with temperature. Specifically, as \(T\) approaches very large values, the potential barrier height diminishes, resulting in a mean first passage time from the AdS plasma phase to the large black hole phase that is proportional to the prefactor raised to the power of three: \(\big<t_{mfp}\big> \propto \frac{\eta}{T^3}\).

The interplay between the MFPT prefactor and the mean transit time is well-defined, with both quantities intricately linked to the curvature of the generalized free energy landscape and temperature. By analyzing black hole time series data, it is possible to experimentally measure the mean transit time and subsequently estimate the MFPT prefactor. Additionally, at elevated temperatures, an inverse relationship between the mean transit time, MFPT, and temperature becomes evident, revealing a direct correlation with the power of temperature.

\subsubsection{RNAdS black hole phase transition}
The relationship between the mean transit time and the prefactor of MFPT remains consistent with equation \eqref{HPprefactor transit time} for the RNAdS black hole phase transition. As the temperature approaches the upper limit, the scaling relationship between the mean transit time and the prefactor will be straightforwardly described.
\begin{figure}[ht]
    \centering
    \subfigure[]{
    \includegraphics[width=0.4\textwidth]{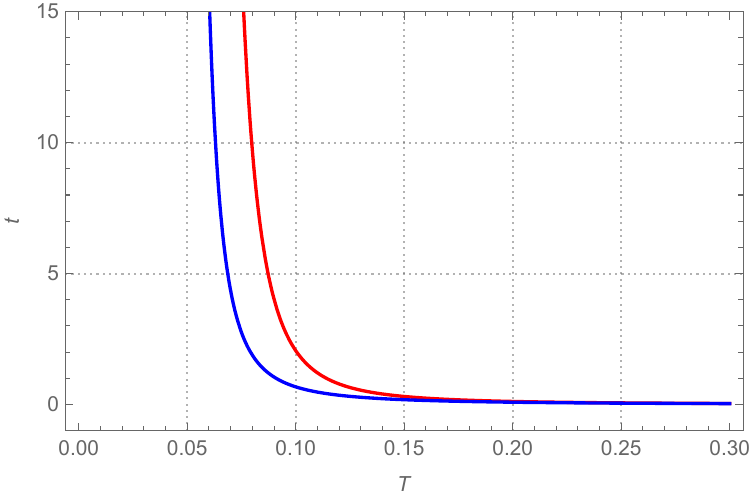}
    }
    \subfigure[]{
    \includegraphics[width=0.4\textwidth]{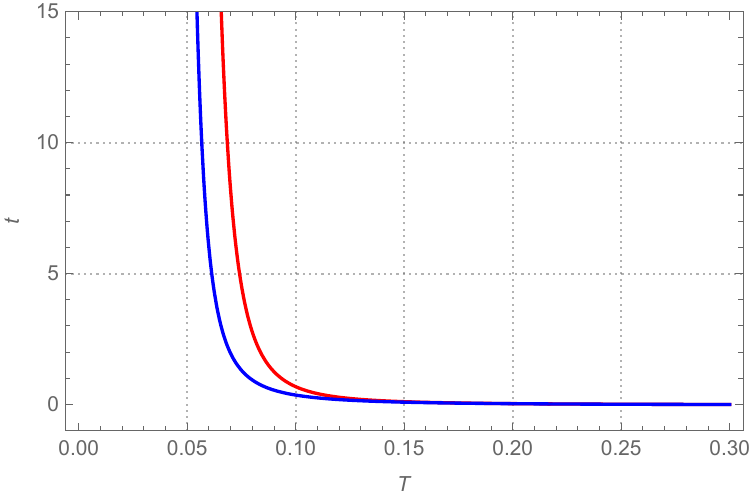}
    }
    \caption{The approximate results of MFPT for the RNAdS black hole compared with the numerical results, where the red lines represent the approximate results and the blue lines represent the numerical results. (a)$Q=0.01 ,L^2=100$ 
 (b)$Q=0.1, L^2=100$ }
    \label{fig:4RNmfpt}
\end{figure}

In equation \eqref{MFPTjs}, it is specified that the MFPT is governed by the generalized free energy landscape of the RNAdS black hole. This landscape determines specific radii \( r_A \), \( r_m \), and \( r_B \) associated with the black hole, which are the roots of the quadratic equation obtained from the extremization of the free energy \(\frac{\partial G(r)}{\partial r} = 0\). Consequently, all relevant information about the MFPT is encoded in these radii, including the contribution of \(\Delta G\), which is also a function of these radii. 

Under the condition where the temperature approaches its maximum value \(T_{\text{max}}\), the MFPT exhibits a relationship of the form:
\begin{equation}
    \big<t_{mfp}\big> \propto \frac{\eta (r_m-r_A)^2}{2T}
\end{equation}

As depicted in Figure \ref{fig:4RNmfpt}, this approximation remains valid for the RNAdS black hole phase transition.

Although the direct dependence on temperature is not explicitly evident in this expression, it is clear that as the charge \(Q\) approaches zero, this expression converges to the case of the Hawking-Page phase transition. This observation indicates that the mean transit time of RNAdS black holes exhibits behavior similar to that of the Hawking-Page phase transition, albeit with more pronounced characteristics.
\begin{equation}\label{tab p}
    \big<t_{AB}\big> \propto \frac{\eta}{\omega'^2} \propto \frac{\eta}{(k \omega)^2}
\end{equation}

Recalling equation \eqref{w'}, as \( r_B \gg r_A \), the parameter \( k \) approaches one. In this case, equation \eqref{tab p} becomes proportional to \(\eta/T\). As the temperature approaches its maximum value \( T_{\text{max}} \), this result becomes increasingly accurate.

\subsection{Landscape impact on approximation methods}
Upon revisiting the derivation of the mean transit time and probability distributions, the process can be outlined as follows. A simplified approach is used to describe the time it takes for a particle to traverse from one basin to another across an intervening potential barrier within a potential field, which is known as the transit time. When random forces and friction are explicitly considered, the curvature is replaced by the square of the curvature divided by the friction coefficient. This adjustment provides a comprehensive description of the transit time. This methodology is derived from reference \cite{Makarov:2010}, which typically focuses on symmetric potential energy scenarios.

However, when studying black hole phase transitions, it is observed that the generalized free energy landscape is significantly influenced by temperature. This requires consideration beyond symmetric landscapes and necessitates an examination of the landscape under various conditions, particularly the impact of temperature.

Attention is drawn to equation \eqref{jifenHPgw} in Appendix \ref{appendix A}, which represents the expression for transit time under the simplified model, as previously mentioned. Recalling the appendix, an approximation was made based on the shape of the landscape, leading to equation \eqref{jifenHPgw1}. We now understand that although the appendix focused on the high-temperature case, the essence lies in the irregularity of the landscape due to the significant influence of temperature. High-temperature examples were prioritized because, as long as the temperature is slightly above the critical temperature \( T_{HP} \), the landscape is predominantly influenced by the right half of the landscape. This situation is relatively common throughout the entire temperature range, from \( T_{HP} \) to infinity or \( T_{max} \).

However, the mean transit time or probability distribution at other temperatures can also be obtained using the same process, as there is no fundamental difference between them; it is the asymmetry of the landscape that causes some quantitative differences. Thus, one can still follow the above process to derive the mean transit time and probability distribution under different temperature conditions.

In the next section, we will specifically discuss the mean transit time and probability distribution as influenced by the effect of temperature on the landscape, i.e., under various temperature conditions.

\section{Different landscapes lead different approximation schemes and behaviors}\label{section 6}
\subsection{Temperature dependence of transit time of Hawking-Page phase transition}
In this subsection, we will examine the behavior of the transit time for the Hawking-Page phase transition.

\subsubsection{High temperature limit}

\begin{figure}[ht]
    \centering
    \includegraphics[width=0.4\textwidth]{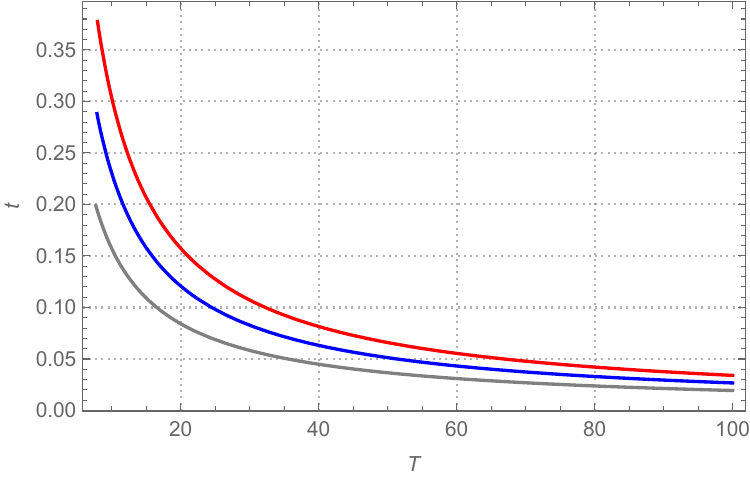} 
    \caption{The mean transit Hawking-Page phase transition from probability average equation \eqref{PAB HP}. The gray,blue and red curves represent L = 10, 100, and 1000, respectively.}
    \label{fig:transit time 10 100 1000}
\end{figure}
From Figure \ref{fig:transit time 10 100 1000}, it is evident that the mean transit time increases with \(L\) at a fixed temperature. This observation is consistent with the integral expression given in equation \eqref{tabjf}, which indicates that as \(L\) increases, the integral expression for \(t\) also increases. Consequently, the statistical average of this value should follow this trend. Analyzing equation \eqref{HPtransit time} reveals the underlying mechanism: the curvature of the generalized free energy landscape and the height of the barrier intuitively explain these behaviors.

To derive a simplified expression for the mean transit time, we perform a Taylor expansion on the minor term \(\frac{T_{\text{min}}^2}{T^2}\) in equations \eqref{eq:rm} and \eqref{eq:rb}. This involves incorporating the substitutions \(r_B\), \(\omega = \sqrt{2\pi T}\), \(T_{\text{min}} = \frac{\sqrt{3}}{2\pi L}\), and the relation \(G^{\neq} = \frac{1}{2}\omega^2(r_B - r_m)^2\).
\begin{equation}\label{HPmfpt1}
   \big<t_{AB}\big>= \frac{\eta ln[\frac{64 \pi^3 T^2 L^4}{9}]}{4 \pi T}
\end{equation}
According to equation \eqref{HPmfpt1}, the curvature is primarily determined by the temperature, while the barrier height \( G^{\neq} \) increases with \( L \). This relationship provides insight into the underlying physical significance of the observed behavior. As previously highlighted, the curvature at the top of the barrier serves as a physical measure of the instability of the intermediate transition state. Consequently, an increase in temperature leads to a higher curvature, indicating that the intermediate transition state black holes become increasingly unstable.

The magnitude of the curvature is directly related to the instability of the intermediate transition state, and this relationship can be analogized to classical models. In the limit where the curvature approaches zero, it resembles a particle at the top of an infinitely flat barrier. In such a scenario, even if the particle is subjected to a random force causing it to deviate from its initial position, the gradient force will be insufficient to rapidly drive the particle down from the barrier top to the basin. Conversely, if the curvature approaches infinity, the situation is akin to a particle at the top of an extremely sharp barrier. In this case, even a slight deviation from its initial position due to a random force will result in a strong gradient force that swiftly drives the particle towards the basin.

The increase in the AdS radius \( L \), which leads to a higher mean transit time, can be understood through the following physical considerations. Analogous to classical models, a particle needs sufficient energy to cross the potential barrier when transitioning from basin A to basin B. Additionally, from the perspective of the intermediate transition state, the height of the barrier reflects the distance between the transition state and other stable states in terms of order parameters or reaction coordinates. Hence, a higher barrier implies a longer mean transit time. As \( L \) increases, it also increases the height of the barrier.

\begin{figure}[!ht]
    \centering
    \subfigure[]{
   \includegraphics[width=0.4\textwidth]{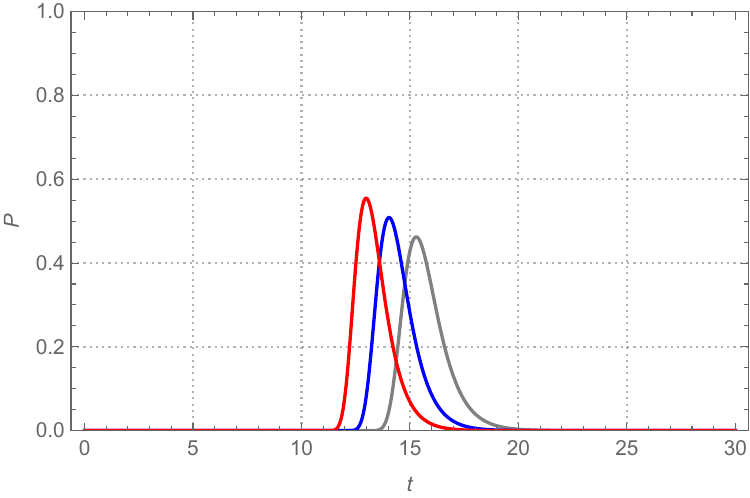} \label{fig:3PHPT}
    }
    \subfigure[]{
    \includegraphics[width=0.4\textwidth]{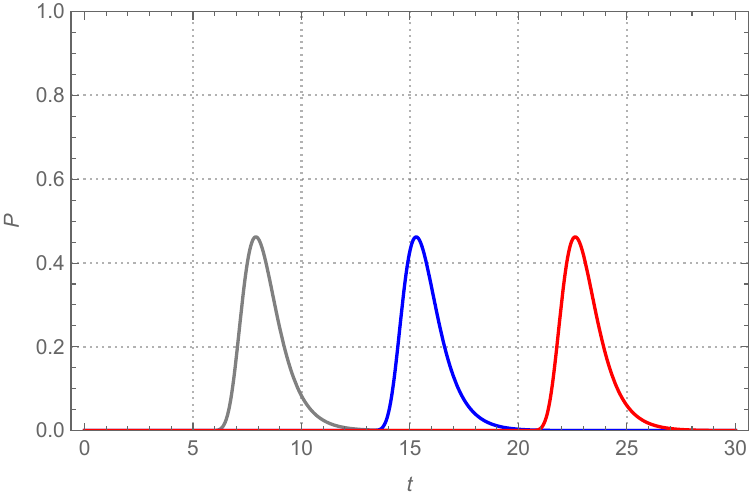} \label{fig:3PHP 10 100 1000}
    }
     \caption{The probability distribution of transit time of the Hawking-Page phase transition from equation \eqref{PAB HP}. (a)The gray,blue and red curves represent temperature T = 0.1, 0.11, and 0.12, respectively.L=100 (b)The gray,blue and red curves represent L = 10, 100, and 1000, respectively.T=0.1.}
\end{figure}
The probability distribution of transit time, as illustrated in Figure \ref{fig:3PHPT}, shifts to the left along the time axis with increasing temperature, indicating a faster rate of the Hawking-Page phase transition. The distribution becomes more peaked and narrower at higher temperatures. At lower temperatures, the distribution broadens, leading to greater fluctuations in transit times.

From the perspective of the free energy landscape, the curvature parameter \(\omega\) increases with temperature, which naturally leads to a reduction in transit time for phase transitions. Higher temperatures also contribute to this effect by enabling the acquisition of sufficient energy in the initial equilibrium state to more rapidly overcome the barrier.

In Figure \ref{fig:3PHP 10 100 1000}, an increase in the AdS radius \(L\) results in the transit time probability distribution shifting to the right along the time axis. This indicates that as \(L\) increases, or equivalently as the cosmological constant decreases, the phase transition process slows down.

Overall, the probability distribution of the transit time adheres to a consistent principle: as the mean transit time decreases, the probability distribution becomes more concentrated around the mean value. This behavior is primarily governed by the curvature, which reflects the stability of the intermediate transition state black hole. If the intermediate transition state is highly unstable, even minor thermal fluctuations can prompt a transition to a stable state. Consequently, thermal fluctuations that can induce a transit event occur within a narrow range, resulting in the transit time being more concentrated around the mean value.

\subsubsection{Low temperature correction}
Since the asymptotic results were derived under the high-temperature approximation, they may not accurately describe low-temperature behavior. Nevertheless, these results can still be used to qualitatively explain certain phenomena.

\begin{figure}[ht]
    \centering
   
    \subfigure[]{
    \includegraphics[width=0.4\textwidth]{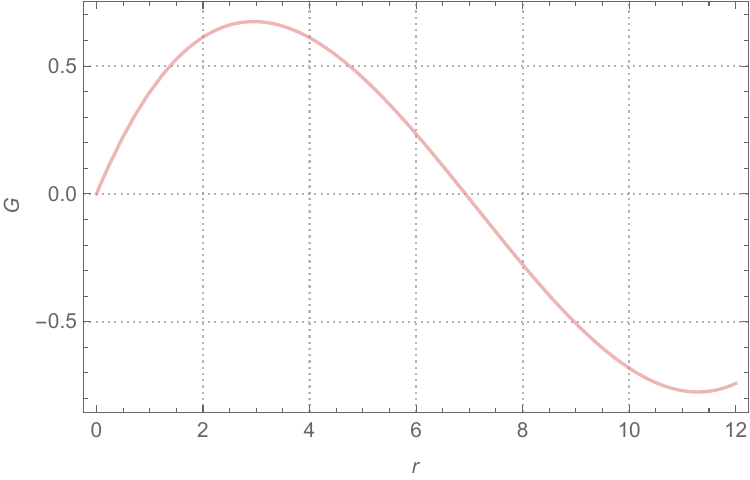} \label{fig:landscape right}
    }
    \subfigure[]{
    \includegraphics[width=0.4\textwidth]{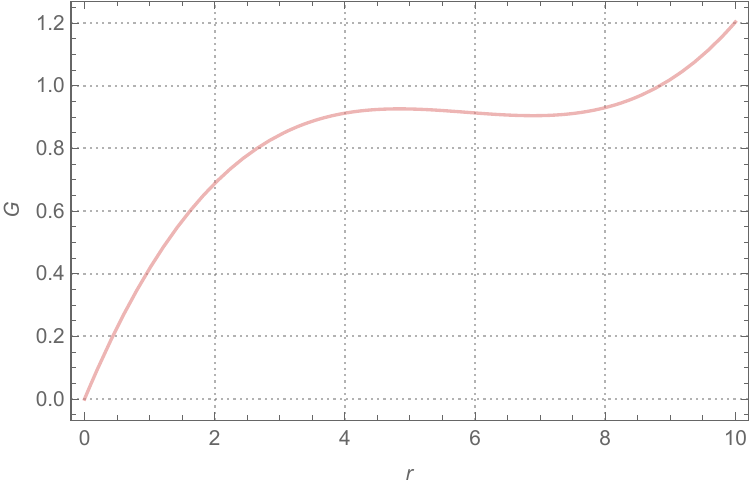 } \label{fig:landscape left}
    }
     \caption{The generalized free energy landscape showing different geometric forms at different temperatures}
\end{figure}
Figures \ref{fig:landscape right} and \ref{fig:landscape left} illustrate variations in the barrier height between the peak at \(r_m\) and \(r_B\), with the barrier being higher on the right in the former and on the left in the latter. As the temperature approaches its extreme values, the difference in barrier heights increases, leading to the observation in the Appendix that, at high temperatures, the barrier resembles only the right half of a parabola.

At low temperatures, \(r_B\) approximates \(r_m\), making the mean transit time's dependence on the barrier height more influenced by the height between \(r_A\) and \(r_m\) rather than between \(r_m\) and \(r_B\), which is the opposite of the high-temperature scenario. Consequently, the free energy barrier affecting the mean transit time should be reassessed as follows:
\begin{equation}
    G^{\neq}_{low} = \frac{1}{2}\omega^2r_m^2
\end{equation}
From equation \eqref{jifenHPgw},it is actually the term $r_m - r_A$ that approaches zero, while the other term is retained. Following the same procedure, one can deduce that the potential barrier height is given by
\begin{equation}\label{HPtransit time low}
    \big<t_{AB}\big>_{low}=\frac{\eta ln[4\beta G^{\neq}_{low}]}{2\omega^2}
\end{equation}
However, it is important to note that the curvature parameter \(\omega\) is still based on the high-temperature approximation. The refinement process involves approximating the free energy as a quadratic function around \(r_m\), achieved by taking the second derivative of the free energy with respect to \(r\). (It is worth noting that this equation reverts to \(\omega = \sqrt{2\pi T}\) in the high-temperature limit.)
\begin{equation}\label{qvlv1}
\omega=\sqrt{2\pi T-\frac{3 r_m}{L^2}}
\end{equation}
Combining the equation for $T_{min}=\frac{\sqrt{3}}{2\pi L}$ and $r_m$, and performing a Taylor expansion of $\epsilon$ in the interval $(T_{min}, T_{min}+\epsilon)$:
\begin{equation}\label{HPtransit time low Talor}
    \big<t_{AB}\big>_{low}=\frac{\pi T_{min}^\frac{3}{2}}{2^{\frac{1}{2}} }\frac{\eta ln[4\beta G^{\neq}_{low}]}{T \epsilon^{\frac{1}{2}} }=\frac{3^{\frac{3}{4}}}{4 \pi^{\frac{1}{2}}} \frac{\eta ln[4\beta G^{\neq}_{low}]}{T L^{\frac{1}{2}} \epsilon^{\frac{1}{2}}}
\end{equation}

\begin{figure}[ht]
    \centering
    \includegraphics[width=0.4\textwidth]{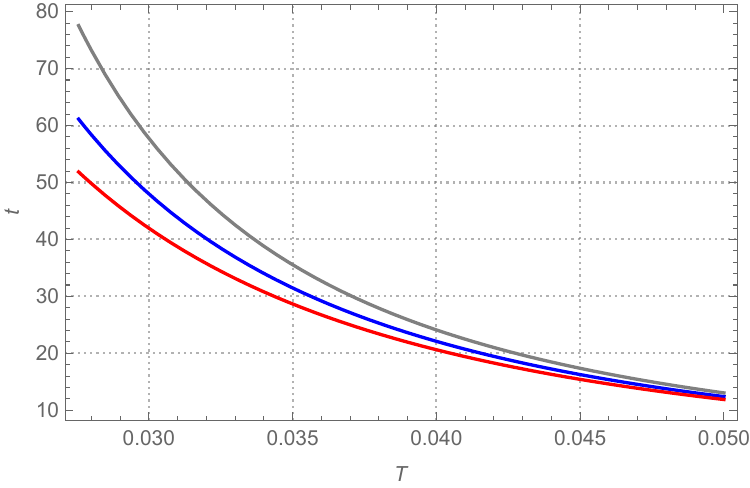}
    \caption{The mean transit time behavior at low temperatures in the case of the Hawking-Page transition.The gray, blue and red lines represent the cases AdS curvature radius $L= 15,20,35$, respectively.}
    \label{t low L10}
\end{figure}
\begin{figure}[ht]
    \centering
   
    \subfigure[]{
    \includegraphics[width=0.4\textwidth]{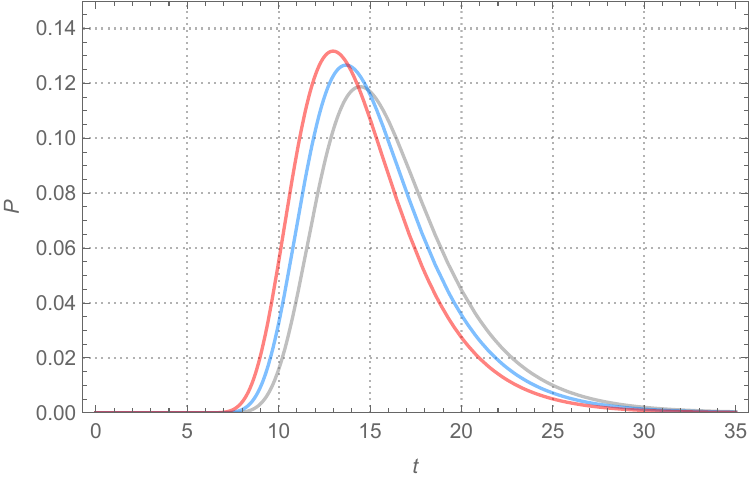} \label{fig:3PHPdw L=35}
    }
    \subfigure[]{
    \includegraphics[width=0.4\textwidth]{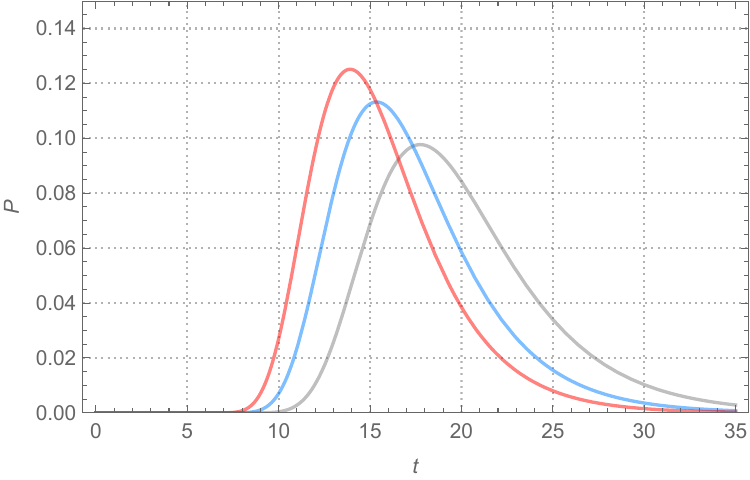 } \label{fig:3PHPdw Tbutong}
    }
     \caption{The probability distribution of transit time after low temperature correction. (a)Gray, blue, and red represent $T=0.027, 0.028, 0.029$with $L=35$. (b)Gray, blue, and red represent $L=15,25,35$with $T=0.027$}\label{fig:3PHdw}
\end{figure}

In Figure \ref{t low L10}, the mean transit time is observed to be a monotonically decreasing function, displaying similarities to the high-temperature behavior but with an adjusted curvature factor. In the region to the right of \(T_{min}\), the curvature approaches an infinitesimal value, causing the mean transit time to decrease rapidly with increasing temperature. Additionally, the mean transit time decreases with an increasing AdS radius \(L\), in contrast to the high-temperature limit results depicted in Figure \ref{fig:transit time 10 100 1000}.

Figure \ref{fig:3PHdw} illustrates a probability distribution that aligns with the previously described trend. Regardless of temperature, the distribution consistently shows that as it shifts leftward (indicating a reduction in mean transit time), the probability becomes more concentrated, signifying reduced fluctuations.

The primary cause of this behavior is the temperature-dependent variation in curvature \(\omega\). According to Equation \eqref{qvlv1}, at low temperatures, curvature is significantly influenced by \(L\). Coupled with Equation \eqref{HPtransit time low Talor}, which shows that the mean transit time is inversely proportional to \(L^{\frac{1}{2}}\), this implies that curvature is proportional to \(L^{\frac{1}{2}}\). In contrast, at high temperatures, curvature depends on temperature, and the mean transit time displays behavior opposite to that observed at low temperatures. This is due to the interaction between barrier height and \(L\). Although the low-temperature barrier also depends on \(L\), the logarithmic relationship between the mean transit time and barrier height suggests that curvature has a more pronounced effect on the mean transit time.

\subsubsection{Behavior near the critical temperature}
The critical temperature for the Hawking-Page phase transition, denoted \(T_{HP} = \frac{1}{\pi L}\), is known as the Hawking-Page critical temperature.
Equation \eqref{jifenHPgw} quantifies the mean transit time, representing the average duration it takes for particles of varying energies to cross a potential barrier from point A to point B. Near the critical temperature, the symmetry in the free energy landscape results in the approximation \( r_B - r_m \approx r_m - r_A \), thereby simplifying the equation.
\begin{equation}
 t=2 arcsinh[\sqrt{\frac{2\pi T}{2E-\frac{1}{8\pi T}}}(r_B-r_m)]/\omega
\end{equation}
By repeating the previous derivation, the mean transit time near the critical temperature is immediately obtained
\begin{equation}\label{HPtransit time THP}
    \big<t_{AB}\big>=\frac{\eta ln[4\beta G^{\neq}]}{\omega^2}
\end{equation}
In the analytical expression, the factor of \(1/2\) vanishes due to the symmetric parabolic approximation of the free energy landscape near the critical temperature. Given the proximity of the Hawking-Page temperature \( T_{HP} = \frac{1}{\pi L} \) and the minimum temperature \( T_{min} = \frac{\sqrt{3}}{2\pi L} \), the mean transit time exhibits behavior similar to the low-temperature scenario but with a doubled value. This similarity introduces increasing error as the temperature nears the critical value. Despite this, the qualitative dependence of the mean transit time on various parameters remains consistent with the low-temperature case, warranting no further detailed discussion here.

The generalized free energy landscape offers valuable insights into these phenomena, as it is highly sensitive to temperature variations. This perspective helps clarify the observed behavior and highlights the landscape's crucial role in understanding the dynamics of mean transit time.

\subsection{Temperature dependence of transit time of RNAdS black hole phase transition}
\subsubsection{High temperature limit}

\begin{figure}[!ht]
    \centering
    \subfigure[]{
    \includegraphics[width=0.4\textwidth]{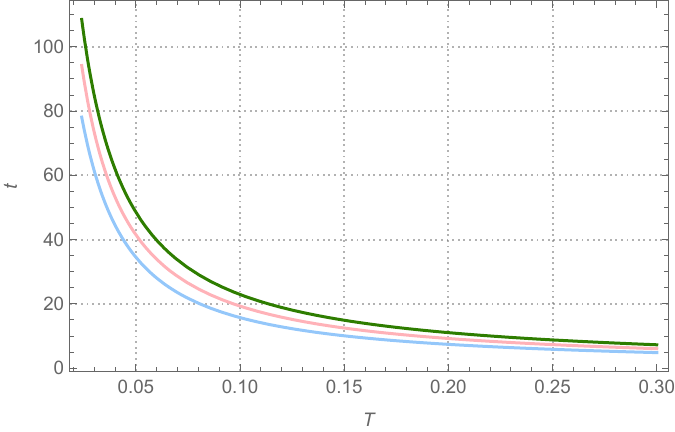}
    }
    \subfigure[]{
    \includegraphics[width=0.4\textwidth]{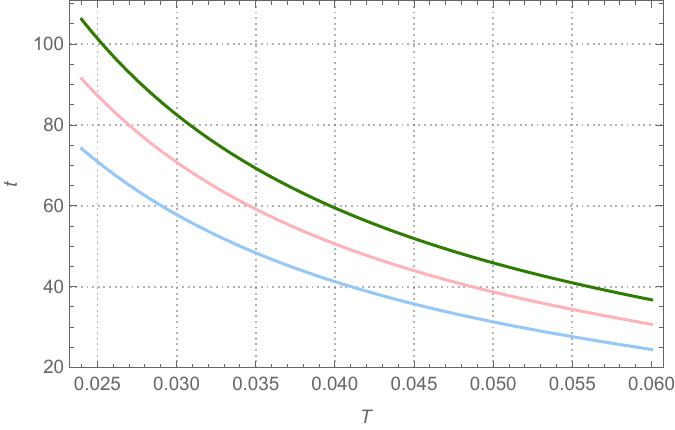}
    }
    
    \caption{The mean transit time as a function of temperature calculated using the probability distribution function $P_{AB}(t)$ under different pressure $P$ (or the cosmological constant $\Lambda$) and charges $Q$. (a) The blue, pink and green lines represent $P=\frac{3}{8\pi}0.001,\frac{3}{8\pi}0.0001,\frac{3}{8\pi}0.00001$ and $Q=0.1$
    (b)The blue, pink and green lines represent $P=\frac{3}{8\pi}0.001,\frac{3}{8\pi}0.0001,\frac{3}{8\pi}0.00001$ and $Q=0.5$}
    \label{fig:RNtpm}
\end{figure}
\begin{figure}[!ht]
    \centering
    \subfigure[]{
    \includegraphics[width=0.4\textwidth]{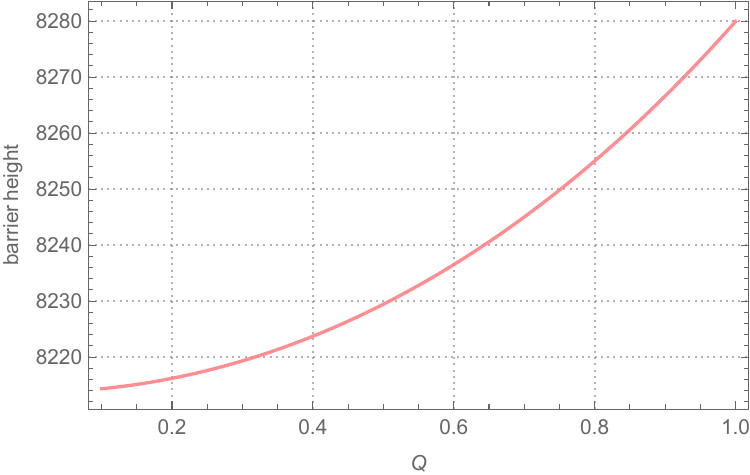}\label{fig:Qheight}
    }
    \subfigure[]{
    \includegraphics[width=0.4\textwidth]{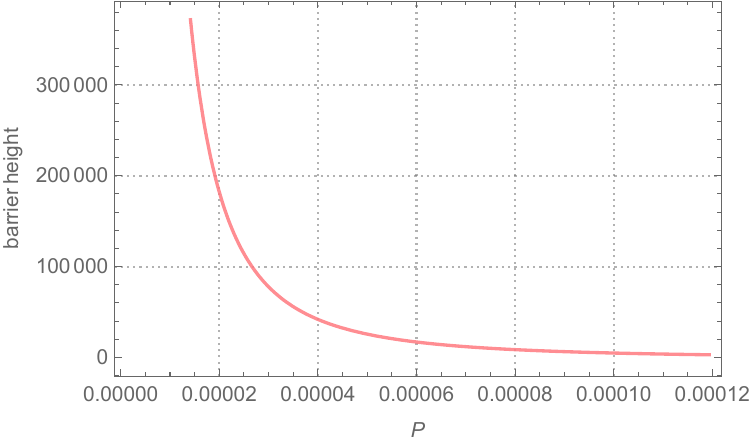}\label{fig:Pheight}
    }
      
    \subfigure[]{
    \includegraphics[width=0.4\textwidth]{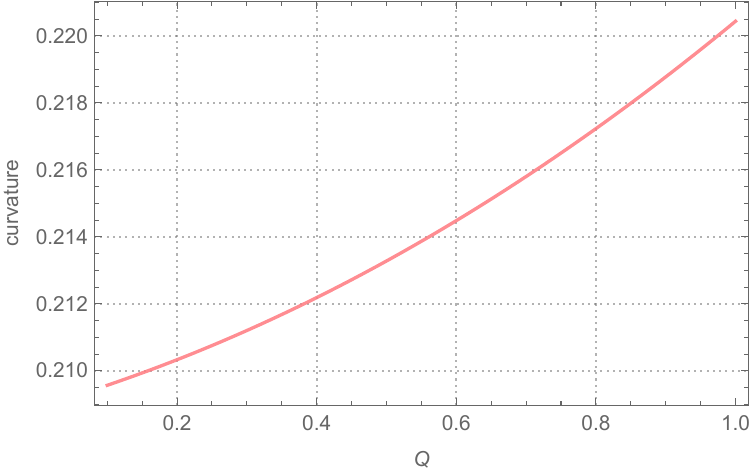}\label{fig:RNcurvature Q}
    }
    \subfigure[]{
    \includegraphics[width=0.4\textwidth]{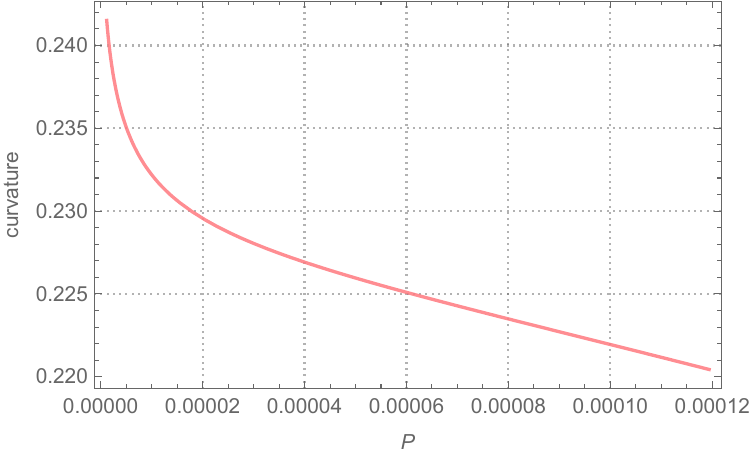}\label{fig:RNcurvature P}
    }
    \caption{The relative barrier height $\beta G^{\neq}$ or curvature as a function of charge $Q$ or pressure $P$( or the cosmological constant $\Lambda$,$P=\frac{3}{8 \pi}\frac{1}{L^2}$,$L=\sqrt{\frac{-3}{\Lambda}}$) at a fixed  temperature.}
    
\end{figure}

    \begin{figure}[!ht]
    \centering
    \subfigure[]{
    \includegraphics[width=0.4\textwidth]{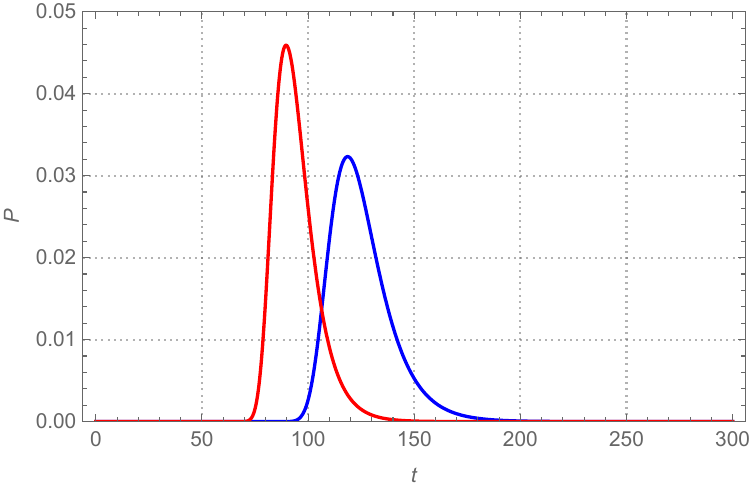}
    }
    \subfigure[]{
    \includegraphics[width=0.4\textwidth]{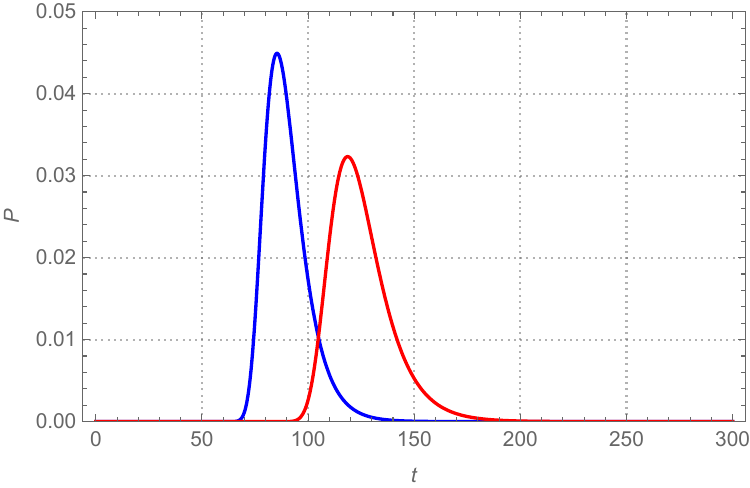}\label{fig:2PRNQbian}
    }
    
    \subfigure[]{
    \includegraphics[width=0.4\textwidth]{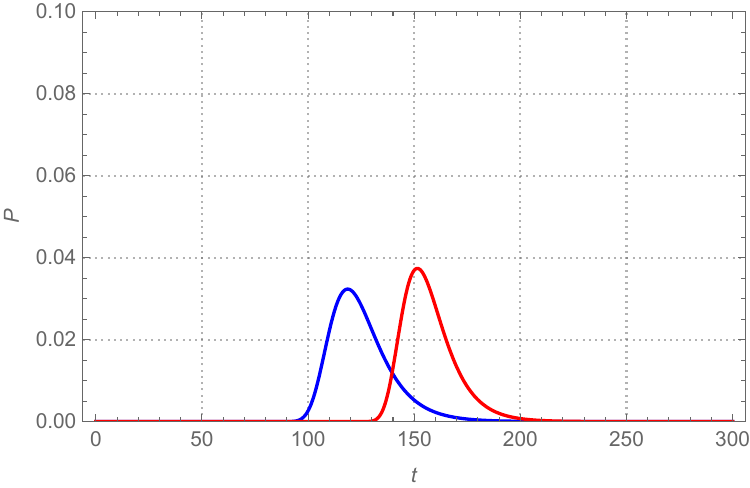}
    }
    \caption{The probability distribution of the transit time for the RNAdS black hole phase transition is influenced by parameters such as temperature $T$, charge $Q$, and pressure $P$( or the cosmological constant $\Lambda$,$P=\frac{3}{8 \pi}\frac{1}{L^2}$,$L=\sqrt{\frac{-3}{\Lambda}}$). (a) The red and blue lines represent $T=0.02$ and $0.015$, respectively.$P=\frac{3}{8\pi}0.001,Q=0.1$. (b) The red and blue lines represent $Q=0.1$ and $2$, respectively.$P=\frac{3}{8\pi}0.001,T=0.015$. (c) The red and blue lines represent $P=\frac{3}{8\pi}0.001$ and $\frac{3}{8\pi}0.0001$, respectively.$T=0.015, Q=0.1$}
\end{figure}
Due to the complexity of analytical solutions for small, large, and intermediate RNAdS black holes, which involve solving a quartic equation, numerical results are more suitable for discussion.

As illustrated in Figure \ref{fig:RNtpm}, the mean transit time decreases with increasing charge \( Q \) and similarly with the parameter \( P = \frac{3}{8\pi} \frac{1}{L^2} \). This behavior can be understood from the analytical expression \eqref{RN tranit time jx}. Unlike the Hawking-Page transitions, RNAdS black holes have a temperature upper limit \( T_{max} \), which is significantly influenced by \( Q \). Notably, the curvature \( \omega' \) near \( T_{max} \) exhibits a temperature dependence distinct from that in Hawking-Page transitions, aligning with the latter only as \( Q \) approaches zero.

Figures \ref{fig:Qheight} and \ref{fig:RNcurvature Q} show that an increase in \( Q \) raises both the barrier height and curvature. The combined effect on the mean transit time, as described by expression \eqref{RN tranit time jx}, is intricate due to the simultaneous increase in both the numerator and the denominator. Nevertheless, numerical results indicate that the primary influence of \( Q \) on the mean transit time is through its effect on the curvature.

In contrast, Figures \ref{fig:Pheight} and \ref{fig:RNcurvature P} demonstrate that an increase in pressure \( P \) (or equivalently, a decrease in the cosmological constant \( \Lambda \)) results in a reduction of both the barrier height and curvature. Consequently, the mean transit time also decreases, primarily due to the rapid reduction in barrier height with increasing \( P \). This observation suggests that, in the high-temperature limit, the mean transit time is predominantly influenced by \( Q \) through its effect on curvature and by \( P \) through its impact on barrier height.

The probability distribution during the RNAdS black hole phase transition follows the established patterns, with temperature and cosmological constant affecting the distribution through their influence on the curvature and barrier height of the generalized free energy landscape. These landscape features, in turn, shape the probability distribution indirectly. Thus, attention is drawn to the charge \( Q \), which emerges as a pivotal parameter influencing the distribution.

From Figure \ref{fig:2PRNQbian}, as the charge \( Q \) increases, the probability distribution shifts to the left along the time axis and becomes more concentrated, resulting in a narrower distribution width. This indicates that as \( Q \) increases, the phase transition rate accelerates, and the fluctuations decrease. This behavior aligns with the earlier conclusion that a decrease in the mean transit time is associated with a reduction in fluctuations.

\subsubsection{Low temperature correction}

In accordance with the same method, the low-temperature situation requires a correction, which is also straightforward. The curvature \(\omega'\) is modified from \(k\sqrt{2\pi T}\) to
\begin{equation}\label{RNqvlvdw}
  \omega'=\sqrt{2\pi T-8 \pi P r_m-\frac{Q^2}{r_m^3}}
\end{equation}
The barrier height is corrected to the following equation
\begin{equation}
   G'^{\neq}_{low} = \frac{1}{2}\omega'^2(r_m-r_A)^2
\end{equation}
So, analytic expressions for low temperatures are obtained
\begin{equation}
  \big<t_{AB}\big>_{low RN}=\frac{\eta ln[4\beta G'^{\neq}_{low}]}{2\omega'^2}
\end{equation}

\begin{figure}[!ht]
    \centering
    \subfigure[]{
   \includegraphics[width=0.4\textwidth]{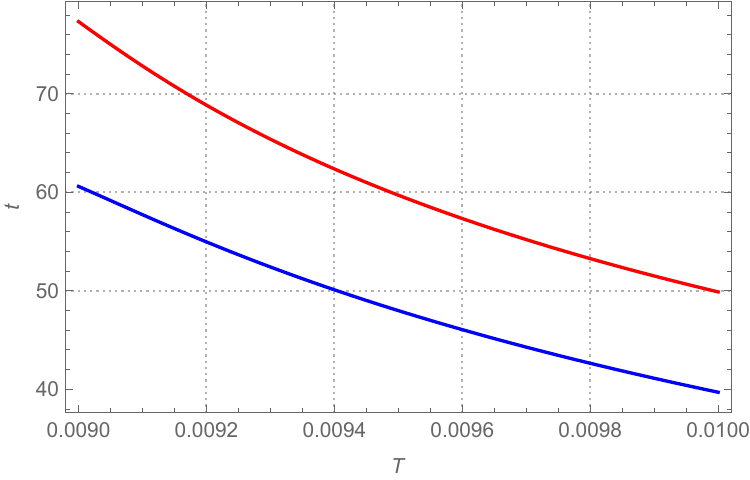}\label{fig:2RNtdwQ}
    }
    \subfigure[]{
    \includegraphics[width=0.4\textwidth]{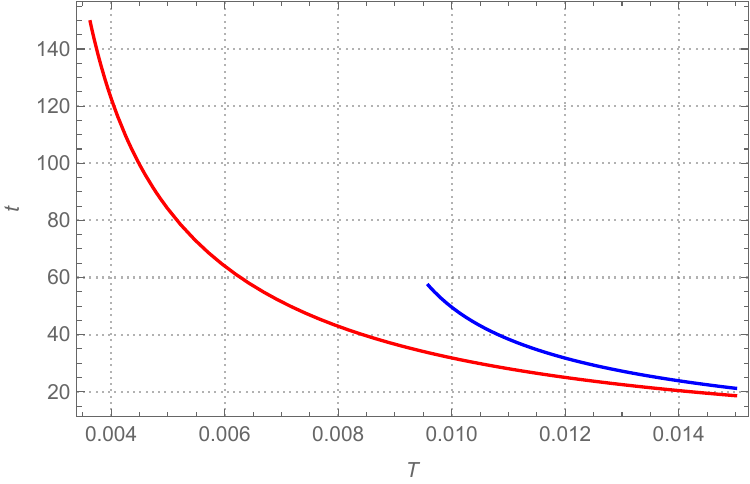}\label{fig:2RNtdw}
    }
    
    \caption{The mean transit time as a function of temperature calculated using the probability distribution function $P_{AB}(t)$ under different pressure $P$ (or the cosmological constant $\Lambda$) and charge $Q$ after low temperature correction. (a) Blue and red line represent $Q=2,0.01$ with $P=\frac{3}{8 \pi}0.001 $, respectively, at low temperatures. (b) Blue and red line represent $P=\frac{3}{8 \pi}0.001 , \frac{3}{8 \pi}0.0001 $ with $Q=0.1$, respectively, at low temperatures} 
\end{figure}
\begin{figure}[!ht]
    \centering
    \subfigure[]{
    \includegraphics[width=0.4\textwidth]{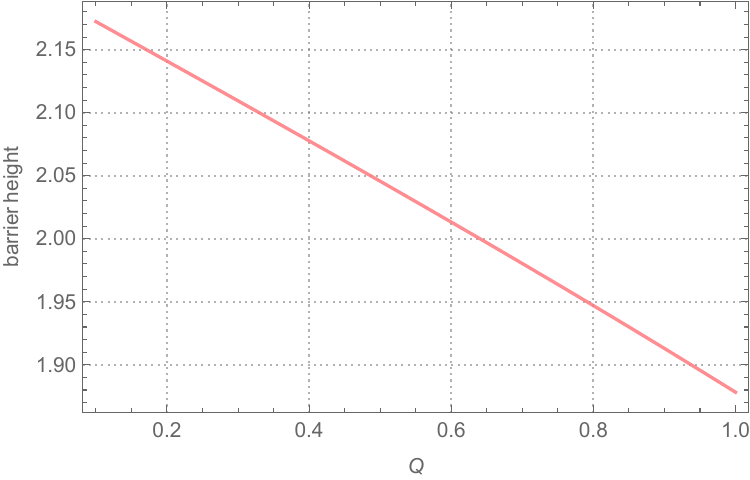}\label{fig:Qheightdw}
    }
    \subfigure[]{
    \includegraphics[width=0.4\textwidth]{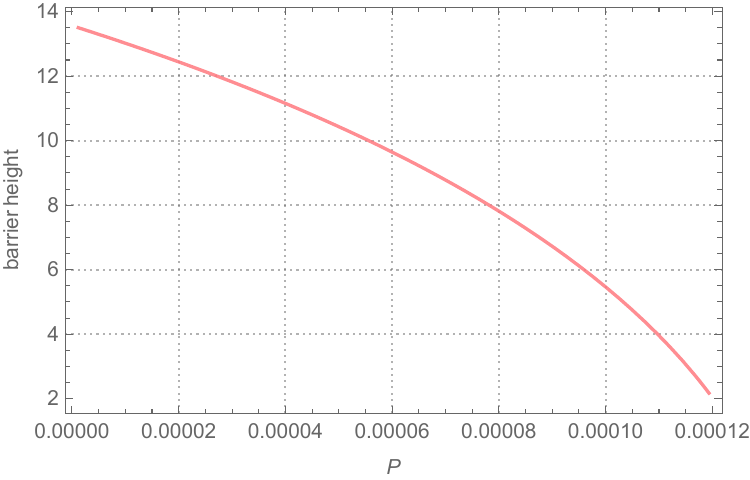}\label{fig:Pheightdw}
    }
    
    \subfigure[]{
    \includegraphics[width=0.4\textwidth]{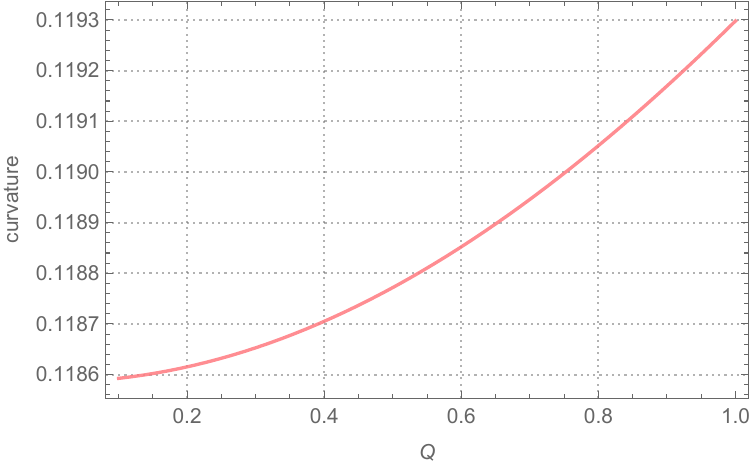}\label{fig:RNcurvature Qdw}
    }
    \subfigure[]{
    \includegraphics[width=0.4\textwidth]{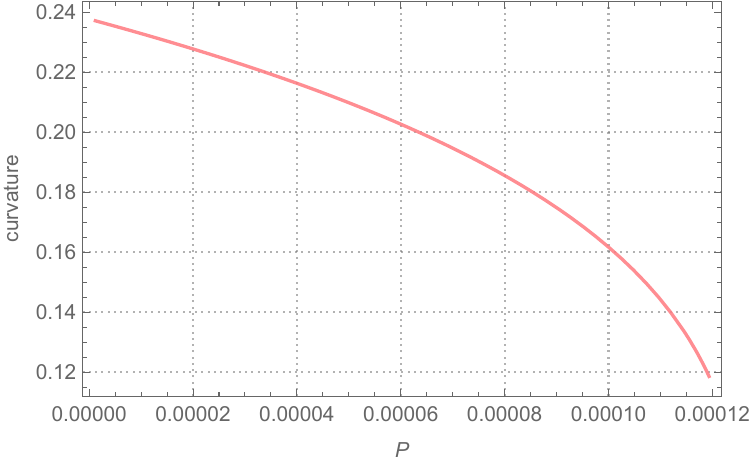}\label{fig:RNcurvature Pdw}
    }
    \caption{The relative barrier height $\beta G^{\neq}$ or curvature $\omega'$ as a function of charge $Q$ or pressure $P$( or the cosmological constant $\Lambda$,$P=\frac{3}{8 \pi}\frac{1}{L^2}$,$L=\sqrt{\frac{-3}{\Lambda}}$) at a fixed low temperature.}
    
\end{figure}

Figure \ref{fig:2RNtdwQ} illustrates that, similar to high-temperature behavior, the mean transit time increases as the charge \( Q \) decreases.

Figure \ref{fig:2RNtdw} reveals that the response of the mean transit time to pressure \( P \) at higher temperatures contrasts with the behavior observed in the Hawking-Page phase transition. This can be approximated from the modified curvature expression. Specifically, as \( P \) decreases, the minimum temperature for phase transitions also decreases, leading to a rapid increase in the mean transit time near the minimum temperature due to the sharp reduction in curvature.

The analysis of mean transit time behavior requires examining the curvature of the unstable state and the barrier height within the generalized free energy landscape. Figures \ref{fig:Qheightdw} and \ref{fig:RNcurvature Qdw} indicate that at low temperatures, the effect of \( Q \) on the mean transit time is clear: an increase in \( Q \) results in a decrease in barrier height and an increase in curvature, thereby reducing the mean transit time.

However, the behavior of \( P \) remains complex. Numerical results indicate that the differing behaviors at low and high temperatures are primarily due to the influence of the barrier height, which is less significant at low temperatures. Figure \ref{fig:Pheightdw} shows minimal changes in the magnitude of the barrier height, while Figure \ref{fig:RNcurvature Pdw} demonstrates that the dependence of the mean transit time on \( P \) is primarily attributed to \( P \)'s effect on curvature, differing from the high-temperature results. This reasoning also applies to the Hawking-Page phase transition, where numerical results offer a more precise understanding of the behavior. The generalized free energy landscape provides a conceptual framework for both qualitative and quantitative analyses.

\subsubsection{Behavior near the critical temperature}

At the critical temperature, the RNAdS black hole phase transition mirrors the Hawking-Page transition closely. In this scenario, the mean transit time is proposed to be twice the magnitude of the low-temperature correction, as detailed in the following equation.
\begin{equation}
  \big<t_{AB}\big>=\frac{\eta ln[4\beta G'^{\neq}_{low}]}{\omega'^2}
\end{equation}
where $\omega'=\sqrt{2\pi T-8 \pi P r_m-\frac{Q^2}{r_m^3}}$.
Based on the low-temperature corrections, it is possible to understand the behaviors at the critical temperature comprehensively. This includes the dependence of the mean transit time on parameters such as temperature \( T \), charge \( Q \), and pressure \( P \); the probability distribution with respect to these parameters; and the reduction in fluctuations as the mean transit time decreases. These behaviors are notably similar to those observed at low temperatures. By analyzing the low-temperature case and then doubling the results, one can effectively grasp the corresponding behaviors at the critical temperature.

\section{Conclusion}
\label{section 7}

In this paper, we investigate the transit time and its probability distribution for both the Hawking-Page transition and the RNAdS black hole phase transition. Our findings reveal that a comprehensive description of the phase transition requires consideration of both the mean transit time and the mean first passage time.

These physical processes in black hole phase transitions can be intuitively understood through the generalized free energy landscape. In this landscape, each stable state corresponds to a basin with a certain probability of remaining in that state or transitioning to another. The mean first passage time (MFPT) represents the average timescale for a stochastic event to occur for the first time due to thermal fluctuations, and it can be described by the Fokker-Planck equation. This timescale is closely related to the mean transit time, which measures the duration required for the system to traverse the barrier from the initial state to the final state, also known as the barrier-crossing time or jump time.

In traditional reaction rate theory, it is generally believed that the MFPT is greater than the mean transit time. However, in the context of black hole phase transitions, we have discovered that the relationship between these timescales can differ due to the free energy landscape’s dependence on temperature. Specifically, the mean transit time is not always less than the MFPT in certain temperature regimes. At lower temperatures, the MFPT is significantly greater than the mean transit time, whereas at higher temperatures, the opposite holds true. Notably, the mean transit time and MFPT are inversely proportional to the cubic and linear powers of temperature, respectively.

This finding highlights the significant impact of temperature on the kinetics of black hole phase transitions, diverging from the expectations of traditional reaction rate theory. The lifetime of a black hole, often calculated based on Hawking radiation evaporation, is typically very long. In contrast, the lifetime of an intermediate black hole, considered as an unstable black hole, can be estimated using the mean transit time. By relating the mean transit time to the prefactor of the mean first passage time, we can estimate the MFPT and thus the lifetime of a stable black hole.

For the low-temperature correction of the mean transit time, we observed that the mean transit time's dependence on the AdS curvature radius \( L \) exhibited behavior opposite to that at high temperatures. This suggests that the temperature-dependent behavior of the mean transit time can be used to more accurately detect the AdS radius \( L \) of black holes. Regardless of whether the temperature is high or low, the mean transit time shows a logarithmic dependence on the barrier height. Thus, by examining the curvature \( \omega \) of the intermediate state black hole on the generalized free energy landscape, we can estimate the mean transit time. Conversely, knowing the mean transit time of the intermediate state black hole allows us to estimate the free energy landscape profile and infer the behavior of small or large black holes. Additionally, through the relationship between the mean transit time and the prefactor of the mean first passage time (MFPT), we can estimate the stable lifetime of black holes.

The mean transit time and mean first passage time, used to describe the black hole phase transition, are related through a friction coefficient \( \eta \). This coefficient is determined by the thermodynamic properties or the interaction strength of the black hole with its environment. Investigating the mean transit time can even reveal properties of the black hole's internal structure. For instance, if experiments are proposed where the black hole is in a thermal environment with varying conditions—such as the same temperature but different optical spectral frequencies determined by the black hole's internal structure—changes in these features might impact the black hole's phase transition. Such experiments could provide insights into the black hole's intrinsic structural properties.

Moreover, the probability distribution of the transit time narrows as the mean transit time decreases, indicating that fluctuations in the transit time also diminish accordingly.

Overall, we conclude that the transit time, as a characteristic timescale describing the kinetics of phase transitions, represents the actual time required for a system to transition from one state to another or to a transition state. Just as Bekenstein’s black hole entropy quantifies the information about the internal state of a black hole in terms of the fundamental units of horizon area, the generalized free energy landscape encapsulates information about the black hole phase transition. By analyzing this landscape—considering its geometric structure, curvature, and barrier height—we can explore the behavior of the transit time and its dependencies on temperature, AdS radius or cosmological constant, and charge. Conversely, understanding the behavior of the transit time (for example, from observational time series of the black hole horizon) may enable us to infer the geometric structure of the free energy landscape and uncover details about the black hole embedded within it.

Furthermore, recognizing the relationship between the mean transit time and the prefactor of the mean first passage time (MFPT) reveals their close connection through the curvature of the free energy landscape. A comprehensive understanding of both the mean transit time and the MFPT provides a complete dynamical picture of black hole phase transitions.

In essence, whether considering high, low, or critical temperatures, an accurate approximation of the generalized free energy landscape can elucidate the mean transit time associated with black hole phase transitions. This perspective remains crucial, as various parameters such as temperature, the AdS radius, and charge inform the geometric structure of the generalized free energy landscape, which in turn dictates the nature of black hole phase transitions.

\section*{Acknowledgments}

TY thanks supports from the National Natural Science Foundation of China GrantNo. 21721003. No.12234019.
\appendix
\section{Appendix:The mean transit time and probability distribution}

\subsection{Hawking-Page phase transition}\label{appendix A}
In this model, the total energy of the particles in the system, denoted $E$, is a conserved quantity that satisfies $E=\frac{p^2}{2m}+V(r)$. The scenario where the initial state of the particles is located within the potential well at point A is first considered. In this case, it can be argued that the initial momentum of the particle follows a random distribution and obeys the Boltzmann distribution law due to thermal equilibrium considerations.

Considering that the particle can move from state $A$ to state $B$ through the barrier, the complete process of each particle moving from A to B is defined as a transit event. Conversely, there are instances where the particle cannot reach point B despite attempting to traverse the barrier. To analyze these events statistically, they are counted and statistically analyzed to obtain the probability of transit event occurrence, denoted as $P_{AB}$.

The flux of particles across the barrier is represented by their momentum (or velocity) flux. Then, $\delta(t-t_{AB})$ is used as a filtering function to select situations that satisfy transit events. Then, the summation and normalization of the flux, after considering the weights, is regarded as the probability distribution. The weights can be determined by the density of states in phase space or, in other words, satisfy the canonical ensemble. The distribution can be obtained from the reference. \cite{Makarov:2010}.

\begin{equation}
 P_{AB}(t)=\int_{p_0}^{\infty}\frac{d p}{2\pi \hbar}(\frac{p}{m})e^{-\beta (\frac{p^2}{2m}+V(r_A))}\delta(t-t_{AB}[\frac{p^2}{2m}+V(r_A)])\bigg/ \int_{p_0}^\infty\frac{d p}{2\pi \hbar}e^{-\beta (\frac{p^2}{2m}+V(r_A))}   
\end{equation}

Then, the integration can be converted into an integration over $E$,
\begin{equation}
P_{AB}(t)=\int_{V_m}^{\infty}\frac{d E}{2\pi \hbar}e^{-\beta E}\delta(t-t_{AB}[E])\bigg /\int_{V_m}^\infty\frac{d E}{2\pi \hbar}e^{-\beta E}
\end{equation}

Another highly significant quantity in this problem is the transit time, which can be obtained from a dynamic perspective. Considering the dynamics of the system, the equation that describes the transit time can be directly derived. Given the conservation of energy $E=\frac{m v^2}{2}+V(r)$, and the relation $v = \frac{dr}{dt}$, the time it takes for a particle to travel from point A to point B can be derived.

Rearranging the equation for energy conservation, we have $\frac{m v^2}{2}=E - V(r)$. Substituting $v = \frac{dr}{dt}$, one can obtain:

\begin{equation}
    t_{AB}=\int_{r_A}^{r_B} \frac{dr}{\sqrt{2(E-V(r))/m}}  
\end{equation}

This equation represents the time it takes for a particle to cross the barrier and move from point A to point B in terms of the potential energy function $V(r)$ and the given energy $E$.

Note that this expression assumes that no external forces or damping effects are present and describes the motion of a classical particle.
In addition, We emphasize again that let us denote $A$ as the starting point and $B$ as the endpoint of the potential barrier. Furthermore, $p_0$ represents the case where the particle has just enough energy to reach the top of the potential barrier from the starting point $A$.
Then, the distribution of transit times and the transit time itself can be utilized to calculate the mean transit time.

Now, the concept of transit time and its distribution is applied to this model. The on-shell free energy is considered to represent the stable phases or states, while the off-shell free energy corresponds to fluctuating phases or states.

To proceed, the metric of the Schwarzschild-AdS black hole and the generalized free energy function, denoted as $G(r)$. Assuming that the black hole is in the AdS space state, one have $M=\frac{r}{2}\left(1+\frac{r^2}{L^2}\right)=0$, and at the same time, $r=0$, which serves as the initial state or phase. As it approaches the limit on the right, the metric approaches that of the AdS space, it is implied that this state is considered as the AdS space state.
\begin{figure}[ht]
    \centering
    \subfigure[]{
     \includegraphics[width=0.4\textwidth]{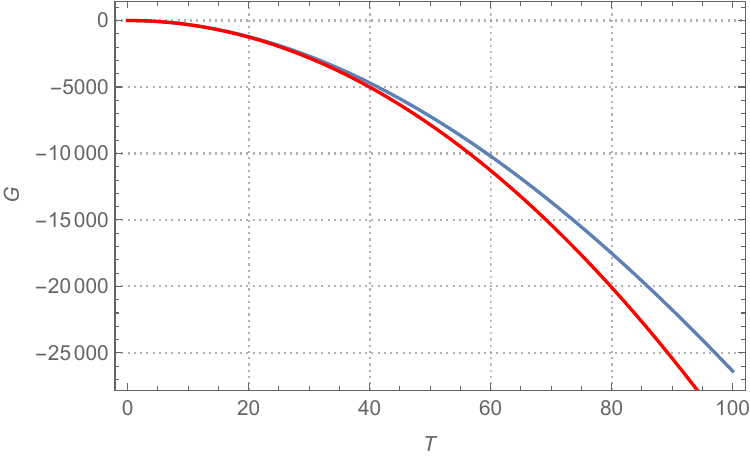}
    }
    \subfigure[]{
     \includegraphics[width=0.4\textwidth]{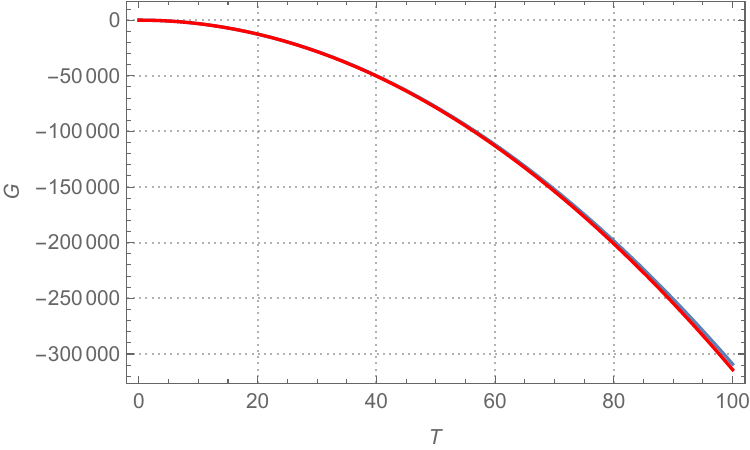}
    }
    \caption{Generalized free energy landscape as a function of horizon size $r$ (a) The blue line is the $G(r)$ ,the red line is approximation $G=G_m-\frac{1}{2}\omega^2 (r-r_m)^2$ when $T=1$,$L=100$.(b)When $T=10$,$L=100$,within certain $r$ limits, the two are almost equal}
    \label{fig:js12}
\end{figure}

Continuing with the discussion, one of the main challenges faced is how to apply the harmonic transition state method to the phase transition of black holes. In the context of black holes, the generalized free energy is viewed as an effective potential energy. Therefore, the transit time $t_{AB}$ should satisfy a functional form similar to that of the harmonic transition state approximation (refer to figure \ref{fig:js12}).
\begin{equation}\label{tabjf}
t_{AB}=\int_{r_{A}}^{r_{B}} \frac{d r}{\sqrt{2(E-G(r))\big/m}}
\end{equation}

Additionally, in the following discussion, it will be assumed, without loss of generality, that the mass $m$ is equal to 1. 
In order to understand the behavior of the transit time, it is necessary to carefully study the dependencies between temperature, the AdS curvature radius $L$ or cosmological constant, and the curvature and barrier height. The method is straightforward: we already know that $r_A = 0$ , and extremize the free energy $G$ with respect to $r$ to get $\frac{\partial G(r) }{\partial r}=\frac{1}{2}+\frac{3r^2}{2L^2}-2\pi T r=0$,
which leads to $r_m$ and $r_B$,

\begin{subequations}
\begin{align}
     r_m=\frac{T}{2\pi T_{min}^2}\bigg(1-\sqrt{1-\frac{T_{min}^2}{T^2}}\bigg) \label{eq:rm} \\
    r_B=\frac{T}{2\pi T_{min}^2}\bigg(1+\sqrt{1-\frac{T_{min}^2}{T^2}}\bigg)\label{eq:rb}
\end{align}
\end{subequations}

Considering how to obtain the distribution of transit times for black hole phase transitions. In order to determine $P_{AB}(t)$, which represents the probability of a black hole undergoing a phase transition and spending a time $t$, we need to establish a fundamental framework. In this model, The generalized free energy can be obtained from the metric, and the topology can be understood. Then, using the canonical ensemble and the ratio of the flux,the probability distribution can be derived.  In this case, the flux is represented by the derivative of the order parameter $r$ with respect to time, denoted as $\dot{r}$. Putting it all together, one obtain the following equation:
\begin{equation}
P_{AB}(t)=\int_{G_m}^{\infty}\frac{d E}{2\pi \hbar}e^{-\beta E}\delta(t-t_{AB}[E])\bigg/\int_{G_m}^\infty\frac{d E}{2\pi \hbar}e^{-\beta E}
\end{equation}
To proceed further, $t$ and $P_{AB}(t)$ are calculated, and the analytical results are obtained. The probability distribution $P_{AB}(t)$ can be expressed as a composition of the $\delta(t-t_{AB}[E])$ function, which can be replaced by $\delta(E-E_{AB})$. As a result, the distribution takes the form:

\begin{equation}
P_{AB}(t)=\int_{G_m}^{\infty}\frac{d E}{2\pi \hbar}\frac{e^{-\beta E}\delta(E-E_{AB})}{\big|\frac{dt}{dE}\big |}\bigg/\int_{G_m}^\infty\frac{d E}{2\pi \hbar}e^{-\beta E}
\end{equation}

Furthermore,one can obtain

\begin{equation}
P_{AB}(t)=\beta e^{\beta G_m} e^{-\beta E_{AB}} \bigg/ \bigg|\frac{d t}{d E}\bigg|_{E=E_{AB}}
\end{equation}

In this case, $E_{AB}$ is the inverse function of $t(E)$.  The free energy $G(r)$ can be approximated as a parabolic potential. There are several reasons why this approximation is reasonable. From equation \eqref{tabjf}, the denominator of the integrand becomes smaller as $G(r)$ decreases, resulting in a smaller contribution to the integration. Furthermore, by examining the equation for $G(r)$.
Therefore, expanding the generalized free energy at the top of the barrier is a reasonable approximation method.

Substituting the approximate expression for $G(r)=G_m-\frac{1}{2}\omega^2 (r-r_m)^2$ into the equation for $t$ derived from \eqref{tabjf}, one obtain:
\begin{equation}\label{jifenHPgw}
 t=arcsinh[\sqrt{\frac{\omega^2}{2E-2G_m}}(r_B-r_m)]-arcsinh[\sqrt{\frac{\omega^2}{2E-2G_m}}(r_A-r_m)]\bigg/\omega
\end{equation}
and $\omega=\sqrt{\frac{2\pi T}{m}}$.
To consider that the $\sqrt{\frac{\omega^2}{2E-2G_m}}(r_A-r_m) \approx 0$,because of $r_A=0$ and $E>G_m$,where $G_m$ is the maximum value when $r$ from zero to $r_B$ and the temperature $T \gg T_{min}$,so the $t$ is that,
\begin{equation}\label{jifenHPgw1}
t=arcsinh[\sqrt{\frac{\omega^2}{2E-2G_m}}(r_B-r_m)]\bigg/\omega
\end{equation}
One step further,one can gain the $E$ as function of $t$,
\begin{equation}
E_{AB}=\frac{G^{\neq}}{sinh^2(\omega t)}+G_m
\end{equation}
and $G^{\neq}=\frac{1}{2}\omega^2(r_B-r_m)^2=\pi T (r_B-\frac{1}{4\pi T})^2$, subscript of $E_{AB}$ is used to represent that $t$ is integral from $r_A$ to $r_B$.
Do all the above and then do some simple calculations, and one can get the following from the reference \cite{Makarov:2010}.
\begin{equation}
P_{AB}(t)=2\omega\beta G^{\neq}  e^{-\frac{\beta G^{\neq}}{sinh^2(\omega t)}}\frac{cosh(\omega t)}{sinh^3(\omega t)}
\end{equation}

\subsection{RNAdS black hole phase transition}\label{appendix B}
RNAdS black hole phase transitions are similar to Hawking-Page phase transitions. A generalized free energy landscape can first be constructed from the line element or metric.
\begin{equation}
ds^{2}=-(1-\frac{2M}{r}+\frac{Q^2}{r^2}+\frac{r^2}{L^2})dt^{2}+(1-\frac{2M}{r}+\frac{Q^2}{r^2}+\frac{r^2}{L^2})^{-1}dr^{2}+r^{2}d\Omega^{2} 
\end{equation}
Then one can get,
\begin{equation}
G=M-TS=\frac{r_+}{2}(1+\frac{r_+^2}{L^2}+\frac{Q^2}{r_+^2})-\pi Tr_+^2
\end{equation}
At the same time, it is known from the Hawking temperature that the phase transition occurs within the following range, which is manifested on the free energy landscape as having two local minima or two stable states.

\begin{equation}
T_{min/max}=\frac{2\sqrt{P}}{\sqrt{\pi}}\frac{1-32\pi P Q^2 +/- \sqrt{1-96\pi P Q^2}}{(1+/- \sqrt{1-96\pi P Q^2})^{\frac{3}{2}}}
\end{equation}
\begin{figure}[!ht]
    \centering
    \includegraphics[width=0.4\textwidth]{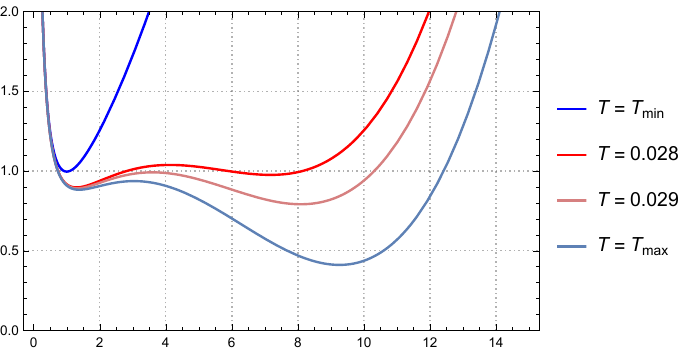}
    \caption{The generalized free energy landscapes against horizon size $r$ when $\frac{1}{L^2}=0.01$.}
    \label{fig:RNL}
\end{figure}
From \ref{fig:RNL},similar to the Hawking-Page phase transition, there are two stable states and a phase transition occurs at a certain temperature under the general free energy landscape.
At this point, a term of $1/r$ provided by the charge appears in the integral,
\begin{equation}
t_{AB}=\int_{r_{A}}^{r_{B}} \frac{d r}{\sqrt{2(E-(\frac{r}{2}(1+\frac{r^2}{L^2}+\frac{Q^2}{r^2})-\pi Tr^2
))\big/m}}
\end{equation}

However, in the phase transition of RNAdS black holes, the same simplification as before cannot be applied, where the latter term was neglected due to $T$ approaching infinity and $r_A$ being equal to 0. In this case, both terms have to be considered from equation \eqref{jifenHPgw}.

One reason to expect only one inverse hyperbolic sine function is that it allows us to express $E$ as a function of $t$. This enables us to obtain an analytical expression for the probability distribution of the transit time. Thus, the equation involving $t$ can be transformed into:
\begin{equation}
 t=arcsinh[\sqrt{\frac{\omega^2}{2E-2G_m}}(r_B-r_m)]\bigg/\omega'
\end{equation}
\begin{equation}
\omega'=\frac{ arcsinh[\sqrt{\frac{\omega^2}{2E-2G_m}}(r_B-r_m)]}{arcsinh[\sqrt{\frac{\omega^2}{2E-2G_m}}(r_B-r_m)]-arcsinh[\sqrt{\frac{\omega^2}{2E-2G_m}}(r_A-r_m)]}\omega=k\omega
\label{w'}
\end{equation}

Indeed, in the equation \ref{w'} involving $t$, $k$ is still a function of $E$. However, if we consider it as a zeroth-order approximation, one can directly obtain the distribution of the RNAdS black hole phase transition. This provides a simplified analytical expression for the probability distribution without taking into account higher-order corrections.
\begin{equation}
P_{AB}(t)=2\omega'\beta G^{\neq}e^{-\frac{\beta G^{\neq}}{sinh^2(\omega' t)}}\frac{cosh(\omega' t)}{sinh^3(\omega' t)}
\end{equation}
Similarly, by applying long-time approximations, the desired result can be obtained from the reference \cite{Makarov:2010},
\begin{equation}
P_{AB}(t)=8\omega' \beta G^{\neq}e^{-2\omega' t -4\beta G^{\neq} e^{-2\omega' t}}
\end{equation}
The analytical expression of mean transit time can be obtained by using the same method as before,
\begin{equation}\label{RNt1}
    \big<t_{AB}\big>=\frac{ln[4\beta G^{\neq}]}{2\omega'}
\end{equation}

The obtained result for the RNAdS black hole phase transition is similar to that of the Hawking-Page phase transition. However, there are modifications in the expression due to the different shape of the free energy landscape. Specifically, the determination of $r_A$ from $\frac{\partial G}{\partial r}=0$ affects both $G^{\neq}$ and subsequently influences other quantities.

Moreover, according to the previous results, the value of $\omega'$ for the RNAdS black hole is multiplied by a parameter $k$, which relates it to the $\omega$ of the Hawking-Page transition.
\begin{figure}[!ht]
    \centering
    \subfigure[]{
    \includegraphics[width=0.4\textwidth]{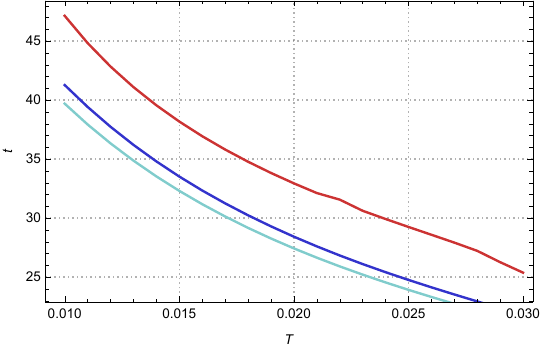}}
     \subfigure[]{
    \includegraphics[width=0.4\textwidth]{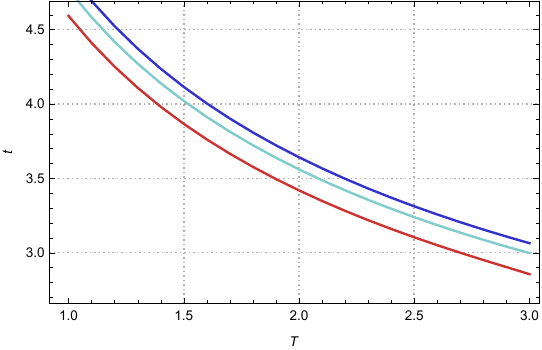}}
    \caption{Mean transit time against temperatures at various estimations with distinct charges $Q$. (a)Blue, dark blue, and red represent analytic expression, probability mean, and ensemble mean, respectively.Q=1 (b)Blue, dark blue, and red represent analytic expression, probability mean, and ensemble mean, respectively.Q=0.01}
    \label{fig:3qt}
\end{figure}

To proceed further, the appropriate value for $E$ needs to be considered when $k$ is treated as a zeroth-order approximation. In order to make this approximation more reasonable, there are common approaches in choosing the value of $E$. One option is to take the minimum value, denoted as $G_m$. Another commonly used approach is to take the average value of $E$, which can be expressed as:
\begin{equation}
\big<E\big>=\int_{G_m}^{\infty}\frac{d E}{2\pi \hbar} Ee^{-\beta E} \bigg/\int_{G_m}^\infty\frac{d E}{2\pi \hbar}e^{-\beta E}=G+\frac{1}{\beta}=G_m+k_BT
\end{equation}
where the $k_B$ is the Boltzmann constant.Using this method, the analytic expression of the mean transit time can also be obtained,
This method is that is the average transition time obtained from the kinetic,then used as the canonical ensemble average
\begin{equation}\label{E mean}
\big<t\big>=\int_{G_m}^{\infty}e^{-\beta E}t_{AB}(E) dE \bigg /\int_{G_m}^{\infty}e^{-\beta E} dE=
\end{equation}
\begin{equation}
\int_{G_m}^{\infty}e^{-\beta E}\int_{r_{A}}^{r_{B}} \frac{d r}{\sqrt{2(E-(\frac{r}{2}(1+\frac{r^2}{L^2}+\frac{Q^2}{r^2})-\pi Tr^2  
))\big/m}}dE\bigg  /\int_{G_m}^{\infty}e^{-\beta E} dE
\end{equation}
When considering the case of $Q=1$ in figure \ref{fig:3qt}, there is still a relatively large error, with the ensemble mean being nearly thirty percent higher than both the analytical expression and probability mean. However, for $Q=0.01$, the ensemble mean is five percent lower than the analytical expression or probability mean. Additionally, as the temperature increases, the disparity between the two decreases. This behavior is particularly pronounced in the Hawking-Page phase transition but is less significant in the RNAdS black hole phase transition.

On the basis of the above analysis, we can conclude that, unlike the Hawking-Page phase transitions, RNAdS phase transitions have not only a minimum temperature but also a maximum temperature. Nonetheless, what they share in common is that near the maximum temperature, the average time obtained through approximation methods becomes increasingly accurate.

\bibliographystyle{plainnat}  
\bibliography{arxiv}  

\begin{thebibliography}{31}
\providecommand{\natexlab}[1]{#1}
\providecommand{\url}[1]{\texttt{#1}}
\expandafter\ifx\csname urlstyle\endcsname\relax
  \providecommand{\doi}[1]{doi: #1}\else
  \providecommand{\doi}{doi: \begingroup \urlstyle{rm}\Url}\fi

\bibitem[Cai et~al.(2013)Cai, Cao, Li, and Yang]{Cai:2013qga}
Rong-Gen Cai, Li-Ming Cao, Li~Li, and Run-Qiu Yang.
\newblock {P-V criticality in the extended phase space of Gauss-Bonnet black holes in AdS space}.
\newblock \emph{JHEP}, 09:\penalty0 005, 2013.
\newblock \doi{10.1007/JHEP09(2013)005}.

\bibitem[Chaudhury and Makarov(2010)]{Makarov:2010}
Srabanti Chaudhury and Dmitrii~E Makarov.
\newblock A harmonic transition state approximation for the duration of reactive events in complex molecular rearrangements.
\newblock \emph{Journal of Chemical Physics}, 133\penalty0 (3):\penalty0 2586, 2010.

\bibitem[Cortes et~al.(1985)Cortes, West, and Lindenberg]{1985On}
Emilio Cortes, Bruce~J. West, and Katja Lindenberg.
\newblock On the generalized langevin equation: Classical and quantum mechanicala).
\newblock \emph{Journal of Chemical Physics}, 82\penalty0 (6):\penalty0 2708--2717, 1985.

\bibitem[Dayyani et~al.(2018)Dayyani, Sheykhi, Dehghani, and Hajkhalili]{Dayyani:2017fuz}
Z.~Dayyani, A.~Sheykhi, M.~H. Dehghani, and S.~Hajkhalili.
\newblock {Critical behavior and phase transition of dilaton black holes with nonlinear electrodynamics}.
\newblock \emph{Eur. Phys. J. C}, 78\penalty0 (2):\penalty0 152, 2018.
\newblock \doi{10.1140/epjc/s10052-018-5623-5}.

\bibitem[Dehyadegari et~al.(2017)Dehyadegari, Sheykhi, and Montakhab]{Dehyadegari:2016nkd}
Amin Dehyadegari, Ahmad Sheykhi, and Afshin Montakhab.
\newblock {Critical behavior and microscopic structure of charged AdS black holes via an alternative phase space}.
\newblock \emph{Phys. Lett. B}, 768:\penalty0 235--240, 2017.
\newblock \doi{10.1016/j.physletb.2017.02.064}.

\bibitem[Dehyadegari et~al.(2019)Dehyadegari, Majhi, Sheykhi, and Montakhab]{Dehyadegari:2018pkb}
Amin Dehyadegari, Bibhas~Ranjan Majhi, Ahmad Sheykhi, and Afshin Montakhab.
\newblock {Universality class of alternative phase space and Van der Waals criticality}.
\newblock \emph{Phys. Lett. B}, 791:\penalty0 30--35, 2019.
\newblock \doi{10.1016/j.physletb.2019.02.026}.

\bibitem[Fernando(2016)]{Fernando:2016sps}
Sharmanthie Fernando.
\newblock {P-V criticality in AdS black holes of massive gravity}.
\newblock \emph{Phys. Rev. D}, 94\penalty0 (12):\penalty0 124049, 2016.
\newblock \doi{10.1103/PhysRevD.94.124049}.

\bibitem[Gubser et~al.(1998)Gubser, Klebanov, and Polyakov]{Gubser:1998bc}
S.~S. Gubser, Igor~R. Klebanov, and Alexander~M. Polyakov.
\newblock {Gauge theory correlators from noncritical string theory}.
\newblock \emph{Phys. Lett. B}, 428:\penalty0 105--114, 1998.
\newblock \doi{10.1016/S0370-2693(98)00377-3}.

\bibitem[Gunasekaran et~al.(2012)Gunasekaran, Kubizňák, and Mann]{gunasekaran2012}
Sundararaman Gunasekaran, David Kubizňák, and Robert~B. Mann.
\newblock Extended phase space thermodynamics for charged and rotating black holes and born-infeld vacuum polarization.
\newblock \emph{J. High Energy Phys.}, 2012\penalty0 (11):\penalty0 1--43, 2012.

\bibitem[Hawking(1974)]{hawking1974}
S.W. Hawking.
\newblock Black hole explosions?
\newblock \emph{Nature}, 248\penalty0 (5443):\penalty0 30--31, 1974.

\bibitem[Hawking(1975)]{Hawking:1975vcx}
S.W. Hawking.
\newblock {Particle Creation by Black Holes}.
\newblock \emph{Commun. Math. Phys.}, 43:\penalty0 199--220, 1975.
\newblock \doi{10.1007/BF02345020}.
\newblock [Erratum: Commun.Math.Phys. 46, 206 (1976)].

\bibitem[J.D.Bekenstein(1973)]{Bekenstein:1973ur}
J.D.Bekenstein.
\newblock {Black holes and entropy}.
\newblock \emph{Phys. Rev. D}, 7:\penalty0 2333--2346, 1973.
\newblock \doi{10.1103/PhysRevD.7.2333}.

\bibitem[J.M.Maldacena(1999)]{J.M.Maldacena:1999}
J.M.Maldacena.
\newblock {The large-N limit of superconformal field theories and supergravity}.
\newblock \emph{Int. J.Theor. Phys.}, 4:\penalty0 38, 1999.
\newblock \doi{1113--1133}.

\bibitem[Kubizňák and Mann(2012)]{kubiznak2012}
D.~Kubizňák and R.~B. Mann.
\newblock P-v criticality of charged ads black holes.
\newblock \emph{J. High Energy Phys.}, 2012\penalty0 (7):\penalty0 1--25, 2012.

\bibitem[Li and Wang(2020)]{Li:2020khm}
Ran Li and Jin Wang.
\newblock Thermodynamics and kinetics of hawking-page phase transition.
\newblock \emph{Phys. Rev. D}, 102\penalty0 (2):\penalty0 024085, 2020.
\newblock \doi{10.1103/PhysRevD.102.024085}.

\bibitem[Li and Wang(2022)]{Li:2022gfe}
Ran Li and Jin Wang.
\newblock {Generalized free energy landscape of a black hole phase transition}.
\newblock \emph{Physical Review D}, 106\penalty0 (10), 2022.
\newblock \doi{https://doi.org/10.1103/PhysRevD.106.106015}.

\bibitem[Mo and Liu(2014)]{Mo:2014qsa}
Jie-Xiong Mo and Wen-Biao Liu.
\newblock {$P-V$ criticality of topological black holes in Lovelock-Born-Infeld gravity}.
\newblock \emph{Eur. Phys. J. C}, 74\penalty0 (4):\penalty0 2836, 2014.
\newblock \doi{10.1140/epjc/s10052-014-2836-0}.

\bibitem[P.Hut(1977)]{P.Hut:1977}
P.Hut.
\newblock {Charged black holes and phase transitions}.
\newblock \emph{Mon. Not. Roy. Astron. Soc.}, 180:\penalty0 379, 1977.

\bibitem[Pollak(1986)]{Pollak1986Theory}
Eli Pollak.
\newblock Theory of activated rate processes: A new derivation of kramers' expression.
\newblock \emph{Journal of Chemical Physics}, 85\penalty0 (2):\penalty0 865--867, 1986.

\bibitem[Rajagopal et~al.(2014)Rajagopal, Kubiz\v{n}\'ak, and Mann]{Rajagopal:2014ewa}
Aruna Rajagopal, David Kubiz\v{n}\'ak, and Robert~B. Mann.
\newblock {Van der Waals black hole}.
\newblock \emph{Phys. Lett. B}, 737:\penalty0 277--279, 2014.
\newblock \doi{10.1016/j.physletb.2014.08.054}.

\bibitem[S.~J.~Yang and Liu(2022)]{2022Kinetics}
S.~W.~Wei S.~J.~Yang, R.~Zhou and Y.~X. Liu.
\newblock Kinetics of a phase transition for a kerr-ads black hole on the free-energy landscape.
\newblock \emph{Physical Review D}, 105\penalty0 (8):\penalty0 084030, 2022.
\newblock arXiv:2105.00491 [gr-qc].

\bibitem[S.W.Hawking and D.N.Page(1983)]{Hawking:1982dh}
S.W.Hawking and D.N.Page.
\newblock {Thermodynamics of Black Holes in anti-De Sitter Space}.
\newblock \emph{Commun. Math. Phys.}, 87:\penalty0 577, 1983.
\newblock \doi{10.1007/BF01208266}.

\bibitem[Wei and Liu(2013)]{Wei:2012ui}
Shao-Wen Wei and Yu-Xiao Liu.
\newblock {Critical phenomena and thermodynamic geometry of charged Gauss-Bonnet AdS black holes}.
\newblock \emph{Phys. Rev. D}, 87\penalty0 (4):\penalty0 044014, 2013.
\newblock \doi{10.1103/PhysRevD.87.044014}.

\bibitem[Wei and Liu(2015)]{2015Insight}
Shao~Wen Wei and Yu~Xiao Liu.
\newblock Insight into the microscopic structure of an ads black hole from a thermodynamical phase transition.
\newblock \emph{Physical Review Letters}, 2015.

\bibitem[Wei et~al.(2019)Wei, Liu, and Mann]{2019Repulsive}
Shao~Wen Wei, Yu~Xiao Liu, and Robert~B Mann.
\newblock Repulsive interactions and universal properties of charged ads black hole microstructures.
\newblock \emph{Physical review letters}, 123\penalty0 (7):\penalty0 071103.1--071103.5, 2019.

\bibitem[Witten(1998{\natexlab{a}})]{Witten:1998qj}
Edward Witten.
\newblock {Anti-de Sitter space and holography}.
\newblock \emph{Adv. Theor. Math. Phys.}, 2:\penalty0 253--291, 1998{\natexlab{a}}.
\newblock \doi{10.4310/ATMP.1998.v2.n2.a2}.

\bibitem[Witten(1998{\natexlab{b}})]{Witten:1998zw}
Edward Witten.
\newblock {Anti-de Sitter space, thermal phase transition, and confinement in gauge theories}.
\newblock \emph{Adv. Theor. Math. Phys.}, 2:\penalty0 505--532, 1998{\natexlab{b}}.
\newblock \doi{10.4310/ATMP.1998.v2.n3.a3}.

\bibitem[Xu et~al.(2015)Xu, Cao, and Hu]{Xu:2015rfa}
Jianfei Xu, Li-Ming Cao, and Ya-Peng Hu.
\newblock {P-V criticality in the extended phase space of black holes in massive gravity}.
\newblock \emph{Phys. Rev. D}, 91\penalty0 (12):\penalty0 124033, 2015.
\newblock \doi{10.1103/PhysRevD.91.124033}.

\bibitem[Yazdikarimi et~al.(2019)Yazdikarimi, Sheykhi, and Dayyani]{Yazdikarimi:2019jux}
H.~Yazdikarimi, A.~Sheykhi, and Z.~Dayyani.
\newblock {Critical behavior of Gauss-Bonnet black holes via an alternative phase space}.
\newblock \emph{Phys. Rev. D}, 99\penalty0 (12):\penalty0 124017, 2019.
\newblock \doi{10.1103/PhysRevD.99.124017}.

\bibitem[Zou et~al.(2014)Zou, Zhang, and Wang]{Zou:2013owa}
De-Cheng Zou, Shao-Jun Zhang, and Bin Wang.
\newblock {Critical behavior of Born-Infeld AdS black holes in the extended phase space thermodynamics}.
\newblock \emph{Phys. Rev. D}, 89\penalty0 (4):\penalty0 044002, 2014.
\newblock \doi{10.1103/PhysRevD.89.044002}.

\bibitem[Zwanzig(2001)]{zwanzig2001nonequilibrium}
R.~Zwanzig.
\newblock \emph{Nonequilibrium Statistical Mechanics}.
\newblock Oxford University Press, 2001.

\end{thebibliography}

\end{document}